\documentclass[letterpaper,12pt]{article}

\usepackage{setspace} 

\usepackage{amsmath}
\usepackage{amsthm}
\usepackage{amssymb}
\usepackage{etoolbox,xparse}
\usepackage[T1]{fontenc}
\usepackage[utf8]{inputenc}
\usepackage{lmodern}
\usepackage{mathtools}
\usepackage[font=footnotesize,labelfont=bf,singlelinecheck=true,labelsep=period]{caption}
\usepackage{xcolor}
\usepackage{titlesec}
\usepackage{enumitem}
\usepackage{array}
\usepackage{booktabs}
\usepackage{tabularx}
\usepackage{csquotes}
\usepackage{needspace}
\usepackage{ragged2e}
\usepackage[english]{babel}
\usepackage[letterpaper,top=1.25in,bottom=1.25in,left=1.25in,right=1.25in,headsep=10pt,marginparsep=0.1cm,marginparwidth=1.9cm]{geometry}
\usepackage{tikz}
\usetikzlibrary{math}
\usetikzlibrary{calc}

\definecolor{mytextcolor}{RGB}{0,0,0}
\usepackage[colorlinks,pdfusetitle,urlcolor=mytextcolor,citecolor=mytextcolor,linkcolor=mytextcolor,bookmarksnumbered,plainpages=false,bookmarks=false,pdfcreator={},pdfproducer={}]{hyperref}
\hypersetup{
  pdftitle={Decomposing the Cost of Complexity Under Risk},
  pdfauthor={Xiaoge Dong and Lasse Mononen},
}
\pdfcatalog{/PageLayout /SinglePage}
\usepackage[nameinlink,capitalise]{cleveref}

\Crefname{axiom}{Axiom}{Axioms}

\usepackage{thmtools}

\declaretheoremstyle[
  spaceabove=5pt plus 2pt minus 3pt,
  spacebelow=5pt plus 2pt minus 3pt,
  headfont=\bfseries,
  notefont=\bfseries, notebraces={(}{)},
  bodyfont=\normalfont,
  postheadspace=8pt plus 1pt minus 2pt,
  headpunct=,
  headindent=,
]{definition}

\declaretheoremstyle[
  spaceabove=4pt plus 2pt minus 3pt,
  spacebelow=4pt plus 2pt minus 3pt,
  headfont=\bfseries,
  notefont=\bfseries, notebraces={(}{)},
  bodyfont=\normalfont,
  postheadspace=8pt plus 1pt minus 2pt,
  headpunct=,
  headindent=,
]{definitionappendix}

\declaretheorem[numbered=no,name=Definition,style=definition]{definition}
\declaretheorem[numbered=yes,name=Theorem,style=definition]{theoremalt}
\declaretheorem[numbered=yes,name=Lemma,style=definitionappendix,sibling=theoremalt]{lemmaalt}
\declaretheorem[numbered=yes,name=Remark,style=definition]{remark}
\declaretheorem[numbered=yes,name=Corollary,style=definition,sibling=theoremalt]{corollaryalt}
\declaretheorem[numbered=yes,name=Proposition,style=definition,sibling=theoremalt]{propositionalt}
\Crefname{lemmaalt}{Lemma}{Lemmas}
\Crefname{theoremalt}{Theorem}{Theorems}
\Crefname{propositionalt}{Proposition}{Propositions}
\Crefname{corollaryalt}{Corollary}{Corollaries}

\NewDocumentEnvironment{theorem}{ o m }{\IfNoValueTF{#1}{\theoremalt}{\theoremalt[#1]}\label{#2}}{\endtheoremalt}
\NewDocumentEnvironment{proposition}{ o m }{\IfNoValueTF{#1}{\propositionalt}{\propositionalt[#1]}\label{#2}}{\endpropositionalt}
\NewDocumentEnvironment{lemma}{ o m }{\IfNoValueTF{#1}{\lemmaalt}{\lemmaalt[#1]}\label{#2}}{\endlemmaalt}
\NewDocumentEnvironment{corollary}{ o m }{\IfNoValueTF{#1}{\corollaryalt}{\corollaryalt[#1]}\label{#2}}{\endcorollaryalt}

\Crefrangeformat{axiom}{Axioms #3#1#4-#5#2#6}
\Crefrangemultiformat{axiom}{Axioms #3#1#4-#5#2#6}{ and #3#1#4-#5#2#6}{, #3#1#4-#5#2#6}{, and #3#1#4-#5#2#6}
\Crefmultiformat{axiom}{Axioms #2#1#3}{ and #2#1#3}{, #2#1#3}{, and #2#1#3}
\declaretheorem[numbered=yes,name=Axiom,style=definition]{axiom}
\AtBeginEnvironment{axiom}{\Needspace{5\baselineskip}}

\declaretheoremstyle[
  spaceabove=\topsep, spacebelow=\topsep,
  headfont=\bfseries,
  notefont=\bfseries, notebraces={(}{)},
  bodyfont=\normalfont,
  postheadspace=8pt plus 1pt minus 2pt,
  headpunct=,
  headindent=,
]{axiomp}
\theoremstyle{axiomp}
\newtheorem{axiomp}{Axiom}

\DeclareMathOperator{\interior}{int}
\DeclareMathOperator{\supp}{supp}
\newcommand*{\R}{\mathbb{R}}
\newcommand*{\N}{\mathbb{N}}
\newcommand{\ndiag}[1]{\operatorname{NDiag}({X^{#1}})}
\newcommand{\longremovespace}{\!\!\!}
\DeclareMathSymbol{\shortminus}{\mathbin}{AMSa}{"39}

\makeatletter
\let\oldvarepsilon\varepsilon
\let\oldepsilon\epsilon
\def\epsilon{\oldvarepsilon}
\def\varepsilon{\oldepsilon}
\let\emptyset\varnothing
\let\oldvarphi\varphi
\let\oldphi\phi
\def\phi{\oldvarphi}
\def\varphi{\oldphi}
\makeatother

\tikzset{
  hollow node/.style={circle,draw,inner sep=1.5},
  solid node/.style={circle,draw,inner sep=1.5,fill=black},
}
\newcommand{\problettercmd}{p}
\newcommand{\branchwhitecmd}{black}
\newcommand{\prizelettercmd}{z}
\newcommand{\prizelettercmdd}{x}
\newcommand{\lotterynamecmd}{P}
\newcommand{\drawtree}{}
\newcommand{\drawtreee}{}
\newcommand{\drawlotteryname}{}

\usepackage[
    backend=biber,
    style=authoryear,
    natbib=false,
    sorting=nyt,
    url=false,
    doi=false,
    isbn=false,
    eprint=false,
    maxnames=50,
    maxcitenames=2,
    uniquelist=false,
    mergedate=compact
]{biblatex}
\DeclareFieldFormat[article,incollection,unpublished]{title}{#1} 
\renewbibmacro{in:}{} 

\usepackage{xpatch}
\xpatchbibmacro{volume+number+eid}{%
  \setunit*{\adddot}%
}{%
}{}{}
\DeclareFieldFormat[article]{number}{\mkbibparens{#1}}
\DeclareNameAlias{author}{family-given} 
\AtBeginBibliography{%
}

\DeclareNameFormat{labelname:poss}{
  \nameparts{#1}%
  \ifcase\value{uniquename}%
    \usebibmacro{name:family}{\namepartfamily}{\namepartgiven}{\namepartprefix}{\namepartsuffix}%
  \or
    \ifuseprefix
      {\usebibmacro{name:first-last}{\namepartfamily}{\namepartgiveni}{\namepartprefix}{\namepartsuffixi}}
      {\usebibmacro{name:first-last}{\namepartfamily}{\namepartgiveni}{\namepartprefixi}{\namepartsuffixi}}%
  \or
    \usebibmacro{name:first-last}{\namepartfamily}{\namepartgiven}{\namepartprefix}{\namepartsuffix}%
  \fi
  \usebibmacro{name:andothers}%
  \ifnumequal{\value{listcount}}{\value{liststop}}{'s}{}}
\DeclareFieldFormat{shorthand:poss}{%
  \ifnameundef{labelname}{#1's}{#1}}
\DeclareFieldFormat{citetitle:poss}{\mkbibemph{#1}'s}
\DeclareFieldFormat{label:poss}{#1's}
\newrobustcmd*{\posscitealias}{%
  \AtNextCite{%
    \DeclareNameAlias{labelname}{labelname:poss}%
    \DeclareFieldAlias{shorthand}{shorthand:poss}%
    \DeclareFieldAlias{citetitle}{citetitle:poss}%
    \DeclareFieldAlias{label}{label:poss}}}
\newrobustcmd*{\posscite}{%
  \posscitealias%
  \textcite}

\DeclareCiteCommand{\parencite}[\mkbibparens]
  {\usebibmacro{prenote}}
  {\usebibmacro{citeindex}%
    \printtext[bibhyperref]{\usebibmacro{cite}}}
  {\multicitedelim}
  {\usebibmacro{postnote}}

\DeclareCiteCommand{\citeauthor}
  {\boolfalse{citetracker}%
   \boolfalse{pagetracker}%
   \usebibmacro{prenote}}
  {\ifciteindex
     {\indexnames{labelname}}
     {}%
   \printtext[bibhyperref]{\printnames{labelname}}}%
  {\multicitedelim}
  {\usebibmacro{postnote}}

\makeatletter
\let\abx@macro@citeOrig\abx@macro@cite
\renewbibmacro{cite}{%
   \bibhyperref{%
   \let\bibhyperref\relax\relax%
   \abx@macro@citeOrig%
   }%
}
\let\abx@macro@textciteOrig\abx@macro@textcite
\renewbibmacro{textcite}{%
   \bibhyperref{%
   \let\bibhyperref\relax\relax%
   \abx@macro@textciteOrig%
   }%
}%
\makeatother

\setlist[itemize]{topsep=2pt,itemsep=0pt,partopsep=0pt,parsep=0pt}
\setlist[enumerate]{topsep=2pt,itemsep=0pt,partopsep=0pt,parsep=0pt}

\titleformat{\section}{\setlength{\RaggedRightLeftskip}{0pt}\RaggedRight\normalfont\large\bfseries}{\thesection}{1em}{}
\titleformat{\subsection}{\setlength{\RaggedRightLeftskip}{20pt}\RaggedRight\normalfont\normalsize\bfseries}{\thesubsection}{1em}{}{}{}
\titleformat{\subsubsection}{\setlength{\RaggedRightLeftskip}{30pt}\RaggedRight\normalfont\normalsize\bfseries}{\thesubsubsection}{1em}{}

\titlespacing*{\section}{0pt}{2.0ex plus 1ex minus 1.5ex}{1.5ex plus 0.5ex minus 0.9ex}
\titlespacing*{\subsection}{20pt}{1.75ex plus 1ex minus 1.3ex}{1.0ex plus 0.5ex minus 0.7ex}
\titlespacing*{\subsubsection}{30pt}{1.75ex plus 1ex minus 1.3ex}{0.8ex plus 0.5ex minus 0.7ex}

\allowdisplaybreaks[1]

\makeatletter
\apptocmd{\small}{\setlength{\abovedisplayskip}{3pt plus2pt minus2pt}\setlength{\belowdisplayskip}{3pt plus2pt minus2pt}\setlength{\abovedisplayshortskip}{0pt plus 3pt minus 2pt}\setlength{\belowdisplayshortskip}{2pt plus 2pt minus 2pt}}{}{}
\apptocmd{\footnotesize}{\setlength{\abovedisplayskip}{3pt plus2pt minus2pt}\setlength{\belowdisplayskip}{3pt plus2pt minus2pt}\setlength{\abovedisplayshortskip}{0pt plus 3pt minus 2pt}\setlength{\belowdisplayshortskip}{2pt plus 2pt minus 2pt}}{}{}
\apptocmd{\normalsize}{
\abovedisplayskip 2\p@ \@plus2\p@ \@minus2\p@
\belowdisplayskip 5\p@ \@plus2\p@ \@minus2\p@
\abovedisplayshortskip 0\p@ \@plus3\p@ \@minus2\p@
\belowdisplayshortskip 3\p@ \@plus3\p@ \@minus3\p@
}{}{}
\makeatother
\makeatletter
\def\@xfootnote[#1]{%
  \protected@xdef\@thefnmark{#1}%
  \@footnotemark\@footnotetext}
\makeatother

\begin{document}

{\normalfont\center\Large\bfseries Decomposing the Cost of Complexity Under Risk \par}
{\center Xiaoge Dong\footnote[$\dagger$]{Chair of Social \& Economic Psychology, Zeppelin University} \& Lasse Mononen\footnote[$\ddagger$]{Center for Mathematical Economics, Bielefeld University. {Updated \today}
\\
This work was funded by the Deutsche Forschungsgemeinschaft (DFG, German Research Foundation), Project-ID 542609382, a joint bilateral project with the PCI2024-155035-2 (MCIN/AEI, Spanish State Research Agency)
}\par}

\begin{abstract}
\noindent
We study preferences under risk when the probability structure of a lottery can affect value separately from its final prizes. Our first result characterizes perturbed expected utility representation in which lottery value separates into expected utility over prizes and an additive perturbation term depending only on the probability tree. The key behavioral restrictions are branch cancellation, which separates prize evaluation from probability structure, and within-tree dominance, which captures that evaluations within the same probability tree are simple and follow dominance. Strengthening branch cancellation to lottery cancellation, which extends the invariance requirement from common prizes to common sublotteries, characterizes additive separability of the perturbation across branches. We also characterize when the tree dependent perturbation term depends only on the number of distinct possible prizes: under weaker continuity requirements, adding same-support independence yields a support size representation for nondegenerate lotteries, and a certainty equivalent condition extends it to degenerate lotteries. Finally, concavity gives the perturbation term its interpretation as a complexity cost.

\end{abstract}
\newpage
\section{Introduction}\label{sec:introduction:complex}

Expected utility evaluates a lottery by averaging utility over its prizes. Probability affects value only by weighting the prizes, so the probability tree or the number of distinct prizes of the lottery carries no separate value. A growing experimental literature shows that lottery valuation is sensitive to complexity and description \parencite{huck1999,levin2016}, including the splitting of probability mass across nearby or equivalent prizes \parencite{stodder1997,bernheim2020}.

To see the issue, consider a lottery that pays $x$ with probability $p$. Split this branch into two branches paying $x$ and $x-\varepsilon$, each with probability $p/2$. As $\varepsilon$ becomes small, the payoff consequences become arbitrarily close to those of the original lottery, while the probability structure changes from one branch to two. Event-splitting effects show that this change can affect choice. We ask when sensitivity to probability structure can be separated from the valuation of final prizes, and when this dependence reduces to the number of distinct possible prizes.

We study the probability structure dependency with the \textit{perturbed expected utility} representation: lottery $P$ is evaluated by
\[V(P)=\sum_{x} P(x) u(x) - c(P),\]
where $u$ is the utility function and $c(P)$ is an additive perturbation term that only depends on the probability structure of $P$. This representation separates the evaluation of final prizes from the effect of probability structure. We derive it from two key behavioral axioms.

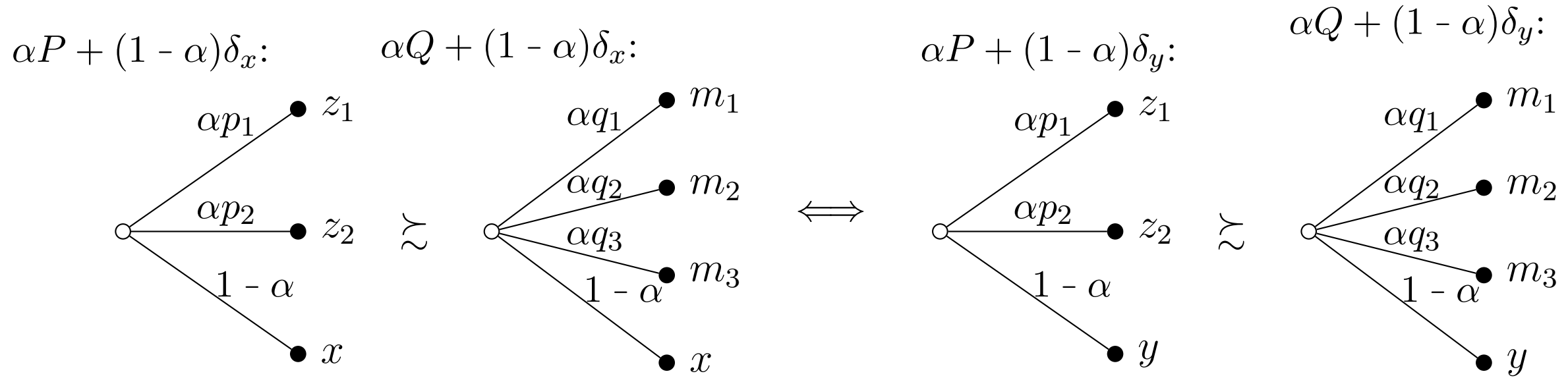
\begin{figure}[t]
\centering
\renewcommand{\drawtree}{%
child[level distance=50pt,sibling distance=35pt,\branchwhitecmd]{node[solid node,\branchwhitecmd,label=right:{$\prizelettercmdd$}]{} edge from parent node[pos=0.6, above, xshift=8pt, yshift=-2pt]{\small$1\shortminus\alpha$}}
child[level distance=50pt,sibling distance=35pt]{node[solid node,label=right:{$\prizelettercmd_2$}](m2){} edge from parent node[pos=0.6, above, xshift=0pt, yshift=-2.5pt]{\small$\problettercmd_2$}}
child[level distance=50pt,sibling distance=35pt]{node[solid node,label=right:{$\prizelettercmd_1$}]{} edge from parent node[pos=0.6, above, xshift=0pt, yshift=2pt]{\small$\problettercmd_1$}};%
}
\renewcommand{\drawtreee}{%
child[level distance=50pt,sibling distance=25pt,\branchwhitecmd]{node[solid node,\branchwhitecmd,label=right:{$\prizelettercmdd$}]{} edge from parent node[pos=0.6, above, xshift=8pt, yshift=-2pt]{\small$1\shortminus\alpha$}}
child[level distance=50pt,sibling distance=25pt]{node[solid node,label=right:{$\prizelettercmd_3$}](m2){} edge from parent node[pos=0.6, above, xshift=0pt, yshift=-2.5pt]{\small$\problettercmd_3$}}
child[level distance=50pt,sibling distance=25pt]{node[solid node,label=right:{$\prizelettercmd_2$}](m2){} edge from parent node[pos=0.6, above, xshift=0pt, yshift=-2.5pt]{\small$\problettercmd_2$}}
child[level distance=50pt,sibling distance=25pt]{node[solid node,label=right:{$\prizelettercmd_1$}]{} edge from parent node[pos=0.6, above, xshift=0pt, yshift=2pt]{\small$\problettercmd_1$}};%
}
\renewcommand{\drawlotteryname}{%
\node at ($(l2)!0.1!(m2)+(0,\lotteryycoord)$) {\small$\lotterynamecmd$:};
}
\begin{tikzpicture}
\tikzmath{\lotteryshift = 3.7;}
\tikzmath{\lotterypairshift = 4.5;}
\tikzmath{\lotteryycoord = 1.8;}

\tikzmath{\lotteryx = 0;}
\renewcommand{\branchwhitecmd}{black}
\renewcommand{\problettercmd}{\alpha p}
\renewcommand{\prizelettercmd}{z}
\renewcommand{\prizelettercmdd}{x}
\coordinate[right](l2) at  (\lotteryx,0);
\node[hollow node] at (\lotteryx,0){}[grow=east]
\drawtree
\renewcommand{\lotterynamecmd}{\alpha P + (1\shortminus\alpha)\delta_x}
\drawlotteryname

\renewcommand{\branchwhitecmd}{black}
\renewcommand{\problettercmd}{\alpha q}
\renewcommand{\prizelettercmd}{m}
\renewcommand{\prizelettercmdd}{x}
\tikzmath{\lotteryx = \lotteryshift;}
\coordinate[right](l2) at  (\lotteryx,0);
\node at ($(l2)!0.4!(m2)$) {\small$\succsim$};
\node[hollow node] at (\lotteryx,0){}[grow=east]
\drawtreee
\renewcommand{\lotterynamecmd}{\alpha Q + (1\shortminus\alpha)\delta_x}
\drawlotteryname

\renewcommand{\drawlotteryname}{%
\node at ($(l2)!0.62!(m2)+(0,\lotteryycoord)$) {\small$\lotterynamecmd$:};
}
\tikzmath{\lotteryx = \lotteryshift + \lotterypairshift;}
\renewcommand{\branchwhitecmd}{}
\renewcommand{\problettercmd}{\alpha p}
\renewcommand{\prizelettercmd}{z}
\renewcommand{\prizelettercmdd}{y}
\coordinate[right](l2) at  (\lotteryx,0);
\node at ($(l2)!0.4!(m2)$) {\small$\iff$};
\node[hollow node] at (\lotteryx,0){}[grow=east]
\drawtree
\renewcommand{\lotterynamecmd}{\alpha P + (1\shortminus\alpha)\delta_y}
\drawlotteryname

\tikzmath{\lotteryx = \lotteryshift + \lotterypairshift + \lotteryshift;}
\renewcommand{\branchwhitecmd}{}
\renewcommand{\problettercmd}{\alpha q}
\renewcommand{\prizelettercmd}{m}
\renewcommand{\prizelettercmdd}{y}
\coordinate[right](l2) at  (\lotteryx,0);
\node at ($(l2)!0.4!(m2)$) {\small$\succsim$};
\node[hollow node] at (\lotteryx,0){}[grow=east]
\drawtreee
\renewcommand{\lotterynamecmd}{\alpha Q + (1\shortminus\alpha)\delta_y}
\drawlotteryname
\end{tikzpicture}
\caption{Branch cancellation. The common branch has probability $1-\alpha$ on both sides. Replacing the common prize $x$ by another common prize $y$ does not affect the comparison.}
\label{fig:weakcindintro}
\end{figure}

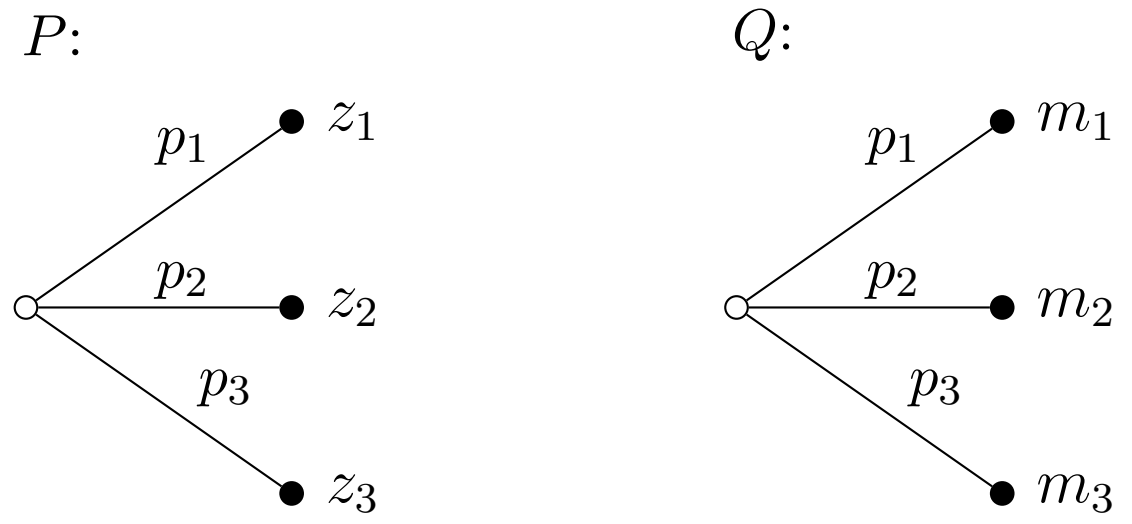
\begin{figure}[t]
\centering
\renewcommand{\drawtree}{%
child[level distance=50pt,sibling distance=35pt,\branchwhitecmd]{node[solid node,\branchwhitecmd,label=right:{$\prizelettercmd_3$}]{} edge from parent node[pos=0.6, above, xshift=8pt, yshift=-2pt]{\small$\problettercmd_3$}}
child[level distance=50pt,sibling distance=35pt]{node[solid node,label=right:{$\prizelettercmd_2$}](m2){} edge from parent node[pos=0.6, above, xshift=0pt, yshift=-2.5pt]{\small$\problettercmd_2$}}
child[level distance=50pt,sibling distance=35pt]{node[solid node,label=right:{$\prizelettercmd_1$}]{} edge from parent node[pos=0.6, above, xshift=0pt, yshift=2pt]{\small$\problettercmd_1$}};%
}
\renewcommand{\drawlotteryname}{%
\node at ($(l2)!0.1!(m2)+(0,\lotteryycoord)$) {\small$\lotterynamecmd$:};
}
\begin{tikzpicture}
\tikzmath{\lotteryshift = 4.7;}
\tikzmath{\lotteryycoord = 1.8;}

\tikzmath{\lotteryx = 0;}
\renewcommand{\branchwhitecmd}{black}
\renewcommand{\problettercmd}{p}
\renewcommand{\prizelettercmd}{z}
\coordinate[right](l2) at  (\lotteryx,0);
\node[hollow node] at (\lotteryx,0){}[grow=east]
\drawtree
\renewcommand{\lotterynamecmd}{P}
\drawlotteryname

\renewcommand{\branchwhitecmd}{black}
\renewcommand{\problettercmd}{p}
\renewcommand{\prizelettercmd}{m}
\tikzmath{\lotteryx = \lotteryshift;}
\coordinate[right](l2) at  (\lotteryx,0);
\node[hollow node] at (\lotteryx,0){}[grow=east]
\drawtree
\renewcommand{\lotterynamecmd}{Q}
\drawlotteryname
\end{tikzpicture}
\caption{Within-tree FOSD. The lotteries $P$ and $Q$ share the same probability tree but different prizes. Then the ranking between $P$ and $Q$ follows FOSD.}
\label{fig:withintreefosdintro}
\end{figure}

The first key axiom, \textit{branch cancellation}, assumes that changing the prize assigned to a common branch in two lotteries leaves their ranking unchanged, provided the branch remains distinct from all others. \Cref{fig:weakcindintro} illustrates the restriction. The second key  axiom is \textit{within probability tree first-order stochastic dominance}, illustrated in \Cref{fig:withintreefosdintro}. This axiom captures the idea that comparing two lotteries is simpler when they share the same induced probability tree, since there are no differences in probability structure to evaluate. Hence, their ranking follows stochastic dominance.

Together with continuity assumptions, these axioms imply the perturbed expected utility representation for lotteries where the probability of each prize is less than 1/2. An additional \textit{restricted common downscaling} axiom extends the representation to all lotteries by requiring that adding a common new prize to two lotteries with the same probability tree cannot reverse their ranking.

The representation accommodates a broad class of perturbations studied in perturbed utility and regularized choice models, including entropy, quadratic, separable convex, and more general symmetric functions of branch probabilities. We provide characterizations for two commonly used special cases. Our first special case characterizes when the perturbation is additively separable across branches. This requires strengthening branch cancellation to \emph{lottery cancellation}, a restricted version of the classical replacement-separability principle for sublotteries discussed by \textcite{machina1989dynamic}. This axiom requires that replacing a common sublottery in two lotteries leaves their ranking unchanged, provided its prizes remain distinct from those in the other branches. Thus, strengthening the familiar cancellation principle from individual prizes to entire sublotteries characterizes additive separability of the perturbation.

We also characterize when the perturbation can be interpreted as a cost of complexity. Under an unbounded utility range, a concavity axiom on preferences is equivalent to concavity of the perturbation function. Concavity makes the perturbation nonnegative and weakly increasing under branch splitting, giving it an interpretation as a cost of complexity.

Next, we consider a discontinuous special case in which the perturbation term depends only on the number of distinct possible prizes. If two lotteries have the same support, they have the same complexity according to this notion. We therefore impose \emph{same-support independence}, which requires the independence axiom on lotteries with the same prizes. We also replace continuity in probabilities with unbounded utilities. Together with branch cancellation and within-tree dominance, these conditions yield that each nondegenerate lottery is evaluated by
\[V(P)=\sum_{x\in\supp P}P(x)u(x)-\kappa(|\supp P|),\]
where $\kappa(|\supp P|)$ is the cost for the support size. A further restriction on certainty equivalents extends this representation to degenerate lotteries. This provides an alternative behavioral foundation for the support size representation studied by \textcite{puri2025}: branch cancellation and within-tree stochastic dominance first separates prize evaluation from the probability tree, while same-support independence restricts the remaining term to depend only on support size. \Cref{tab:roadmap} summarizes our decomposition of the cost of complexity.

\begin{table}[t]
\centering\footnotesize
\caption{Roadmap of the characterizations.}
\label{tab:roadmap}
\begin{tabularx}{\textwidth}{@{}m{0.33\textwidth} X l@{}}
\toprule
\textbf{Behavioral restrictions}
& \textbf{Representation}
& \textbf{Result} \\
\midrule

Branch cancellation + single-prize monotonicity + continuity
&
$\sum_x v(P(x),x)-c(P)$
&
\Cref{prop:perturbedeuwithoutfosd}
\\[2.0ex]

+ Within-tree FOSD
&
$\sum_x P(x)u(x)-c(P)$, if $P(x)\leq\frac12$ for all $x$
&
\Cref{thm:variationalweightedfull}
\\[1ex]

+ Restricted downscaling
&
$\sum_x P(x)u(x)-c(P)$
&
\Cref{thm:variationalweighted}
\\[1ex]

+ Lottery cancellation
&
$\sum_xP(x)u(x)-\sum_x\psi(P(x))+\psi(1)$
&
\Cref{thm:separablevariationalweighted}
\\[1ex]

+ Concavity
&
\begin{tabular}[c]{@{}l@{}}$\sum_xP(x)u(x)-\sum_x\psi(P(x))+\psi(1)$ and\\
\hspace*{1.5em}$\sum_x\psi(P(x))-\psi(1)$ is concave in $P$\end{tabular}
&
\Cref{thm:concave-information}
\\
\midrule

\multicolumn{3}{@{}l}{\textit{Alternative route: support size dependence}}\\
Branch cancellation + within-tree FOSD + unboundedness + same-support independence
&
\vspace{-1.1cm}
\[\begin{dcases*}
\textstyle{\sum_xP(x)u(x)-\kappa(|\supp P|)}, &\text{if $|\supp P|>1$,}\\
\phi(u(x)), &\text{if $P=\delta_x$}
\end{dcases*}\]
&
\Cref{thm:costforsupport}
\\[1ex]

+ Certainty equivalent swap
&
$\sum_xP(x)u(x)-\kappa(|\supp P|)$
&
\Cref{prop:costforsupportfull}
\\[1ex]

+ Concavity
&
$\sum_xP(x)u(x)-\kappa(|\supp P|)$, $\kappa$ weakly increasing
&
\Cref{cor:monotone-support-size-costs}
\\

\bottomrule
\end{tabularx}
\end{table}

We apply the representation to event splitting and deductible insurance. In event-splitting designs, bringing distinct prizes closer together changes expected utility continuously while leaving the perturbation unchanged. When the prizes coincide, however, their branches merge, potentially generating a discrete change in valuation.  Deductible insurance exhibits a similar pattern: Varying the deductible between adjacent loss levels changes final wealth without changing the induced probability tree. When the deductible falls to a loss level, however, that loss and all larger losses yield the same final wealth, causing their branches to merge. Under a preference for simplicity, such mergers weakly reduce complexity costs.

Existing models of complexity under risk typically impose a particular notion of complexity, such as support size \parencite{puri2025}, entropy \parencite{mononen2026preference}, or distance from certainty \parencite{principi2026distance}. Rather than imposing a particular measure of complexity, our characterization leaves the tree dependent component unrestricted and shows what is the behavioral principle giving rise to the complexity cost.

Our characterization also complements the behavioral foundations of perturbed utility in stochastic choice \parencite{fudenberg2015stochastic}. Whereas stochastic-choice data reveal only the optimal probability distribution over alternatives for each menu, we consider the richer preference domain in which all probability distributions can be compared pairwise, and provide behavioral foundations for perturbed utility on this domain. Finally, adding concavity gives the perturbation a complexity-cost interpretation: the general perturbation is nonnegative and weakly increases under branch splitting, while the support size cost is weakly increasing.

The remainder of the paper proceeds as follows. \Cref{sec:relatedliterature:complex} discusses the related literature. \Cref{sec:preliminaries} introduces the setting. \Cref{sec:perturbed-eu} presents the perturbed expected utility characterization. \Cref{sec:extension:complex} gives the support size characterization. \Cref{sec:preference_for_simplicity} characterizes when the perturbation represents a preference for simplicity. \Cref{sec:examples_applications} develops the applications, and \Cref{sec:conclusion:complex} concludes. Proofs are in the appendix.

\subsection{Related Literature}\label{sec:relatedliterature:complex}

This paper contributes to the literature on complexity in choice under risk. Experimental work documents sensitivity to the number of outcomes, probability structure, and lottery description \parencite{mador2000,moffatt2015,bernheim2020,oberholzer2024complexity,puri2025}, while \textcite{goodmanpuri2025} document field demand for simple binary bets even when cheaper dominating alternatives are available.

The closest axiomatic comparisons are \textcite{puri2025,mononen2026preference,principi2026distance}. Our axiomatization for the cost for the size of support complements \citeauthor{puri2025}'s (\citeyear{puri2025}) characterization by decomposing it and offering an alternative foundations for the model by standard axioms and connect the support size representation to the general cost of complexity representation. \textcite{mononen2026preference} relaxes the independence axiom for the event-splitting but maintains that common branches are removed and hence rules out the common-ratio effect. This characterizes entropy-modified and power-weighted expected utility. Our more general perturbation functions accommodate the common-ratio effect. \textcite{principi2026distance} proposes a distance-based complexity index based on the distance to the closest degenerate lottery.

Perturbed utility is closely related to random utility models of stochastic choice. For a fixed set of alternatives and given utility indices, \textcite{hofbauer2002} show that the choice probabilities generated by sufficiently regular additive random utility models can be obtained by maximizing perturbed expected utility with a convex perturbation function. \textcite{fudenberg2015stochastic} provide revealed preference foundations for perturbed utility with an additively separable perturbation using stochastic choice across menus. In their setting, the observed choice probabilities correspond to the optimal probability distribution over alternatives for each menu, leaving preferences over nonoptimal distributions unobserved. By contrast, we take preferences over all probability distributions as primitive, allowing us to impose behavioral restrictions directly on their pairwise comparisons. Our characterization yields a more general perturbed expected utility representation in which the perturbation need not be additively separable across branches.

Branch cancellation is a canonical cancellation axiom in the literature on choice under risk. The underlying idea goes back to the cancellation operation in the editing phase of \textcite{kahneman1979} and appears in the risk literature as branch independence and common branch substitution \parencite{birnbaum2008,lucegamble22008}.\footnote{A related invariance condition appears in the ambiguity literature as weak certainty independence in \textcite{maccheroni2006ambiguity}.}
However, the characterization of the branch cancellation axiom has remained unknown. We fill this gap by showing that it characterizes a representation with a perturbation term depending only on the induced probability tree.

The stronger lottery cancellation axiom extends this invariance requirement from common prizes to common sublotteries. It is a restricted version of the classical replacement separability principle for sublotteries discussed by \textcite{machina1989dynamic}. Related independence conditions also appear in stochastic choice models \parencite{blavatskyy2008stochastic,ke2018rational}. This principle has substantial normative appeal. \textcite{NielsenRehbeck2022} elicit which canonical decision rules subjects want their choices to satisfy and find that $82\%$ choose lottery cancellation. We show that lottery cancellation characterizes additive separability of the perturbation across branches.

Finally, the representation is related to models of process utility and the utility of gambling \parencite{conlisk1993,menestrel2001,diecidue2004,lucegamble12008,lucegamble22008}. The perturbation term captures a component of value attached to the lottery's probability structure rather than its consequences. Its sign determines whether this component acts as a complexity cost or as intrinsic utility from the lottery-resolution process.

\section{Setting}
\label{sec:preliminaries}

We consider a standard setting in choice under risk with monetary prizes. The prize space $X\subseteq\R$ is a nonempty open interval. $\Delta(X)$ is the set of simple lotteries on $X$. Preferences $\succsim$ are defined over lotteries $P\in\Delta(X)$. $\supp P$ denotes the support of the lottery $P$.\footnote{For simple lotteries, $\supp P=\{x\in X\mid P(x)>0\}$.} For $x\in X$, $\delta_x$ denotes the degenerate lottery yielding $x$ with certainty. We endow $\Delta(X)$ with the weak topology.\footnote{For simple lotteries, weak convergence simplifies to convergence of cumulative distribution functions outside the support of the limit lottery: a sequence of lotteries $(P_n)_{n=1}^\infty\subseteq\Delta(X)$ converges weakly to $P\in\Delta(X)$ if, for all $a\in\R$ with $a\notin\supp P$, $\sum_{x\in\supp P_n, x\leq a}P_n(x)\to\sum_{x\in\supp P,x\leq a} P(x)$\ as $n\to\infty.
$}

Mixtures are defined prizewise, and compound lotteries are reduced to single-stage lotteries: for all $\alpha\in[0,1]$, $P,Q\in\Delta(X)$, and $x\in X$, define $\big(\alpha P+(1-\alpha)Q\big)(x)=\alpha P(x)+(1-\alpha)Q(x)$.

We separate a lottery into its induced probability tree and its prize vector, following \textcite{handa1977}. \Cref{fig:withintreefosdintro} illustrates two lotteries with the same induced probability tree but different prize vectors. Formally:

\begin{definition}
\label{def:sameindprobtree}
Lotteries $P$ and $Q$ \emph{have the same induced probability tree} if there exist $n\in\N$, $p=(p_1,\ldots,p_n)\in(0,1]^n$ with $\sum_{i=1}^n p_i=1$, and $(x_i)_{i=1}^n,(y_i)_{i=1}^n\in X^n$ such that $x_i\neq x_j$ and $y_i\neq y_j$ for all $i\neq j$, $P=\sum_{i=1}^n p_i\delta_{x_i}$, and $Q=\sum_{i=1}^n p_i\delta_{y_i}$.
\end{definition}

In \Cref{fig:withintreefosdintro}, if $z_1=z_2$, then the two branches coalesce: $p_1\delta_{z_1}+p_2\delta_{z_1}=(p_1+p_2)\delta_{z_1}$. This creates a discontinuous change in the complexity of the lottery that our models capture.

\section{Perturbed Expected Utility}
\label{sec:perturbed-eu}

We consider an additive perturbation to expected utility that depends only on the induced probability tree.

\begin{definition}
\label{def:information-function}
A function $c:\Delta(X)\to\R$ is a \emph{perturbation function} if $c(P)=c(Q)$ whenever $P$ and $Q$ have the same induced probability tree, and $c(\delta_x)=0$ for all $x\in X$.

A function $c:\Delta(X)\to\R$ is \emph{linearly continuous} if, for all $P,Q\in\Delta(X)$, $c(P)=\lim_{\alpha\to1}c\big(\alpha P+(1-\alpha)Q\big)$.
\end{definition}

\begin{definition}
\label{def:perturbed-eu}
Preferences $\succsim$ admit a \emph{perturbed expected utility representation} if there exist a utility function $u:X\to\R$ and a perturbation function $c:\Delta(X)\to\R$ such that
\[P\succsim Q\iff\longremovespace\sum_{x\in\supp P}\longremovespace P(x)u(x)-c(P)\geq\longremovespace\sum_{x\in\supp Q}\longremovespace Q(x)u(x)-c(Q),\quad\text{ for all $P,Q\in\Delta(X).$}\]
\end{definition}

The expected utility term evaluates prizes, while the perturbation function depends only on the induced probability tree and vanishes for certainty.

We characterize the perturbed expected utility representation in two steps. The first five axioms yield the representation for lotteries where each branch has probability less than $1/2$. For branches with probability greater than $1/2$, the probability and utility are inseparable. Adding \Cref{ax:splittinghalf} the representation to all the lotteries. The first four axioms impose weak order, continuity, and within-tree stochastic dominance.

\begin{axiom}[Weak Order]\label{ax:weakorder:simple}
$\succsim$ is complete and transitive.
\end{axiom}

Continuity is imposed separately on prizes and probabilities. We assume continuity in prizes holding the induced probability tree fixed. This allows event-splitting effects when the induced probability tree changes.\footnote{For lotteries with the same induced probability tree, weak convergence is equivalent to prizewise convergence. In particular, if $(P_n)_{n=1}^\infty$ and $P$ have the same induced probability tree $(p_1,\ldots,p_k)$, then $P_n$ converges weakly to $P$ as $n\to\infty$ iff they can be written as $P_n=\sum_{i=1}^k p_i\delta_{x_i^n}$ and $P=\sum_{i=1}^k p_i\delta_{x_i}$, with $x_i^n\to x_i$ for every $i=1,\ldots,k$.}

\begin{axiom}[Within-Tree Prize Continuity]
\label{ax:weakcontinuityinmoney:simple}
For all lotteries $(P_n)_{n=1}^\infty$, $P$, and $Q$ such that $P_n$ and $P$ have the same induced probability tree for every $n\in\N$ and $P_n$ converges weakly to $P$ as $n\to\infty$, if $P_n\succsim Q$ for all $n\in\N$, then $P\succsim Q$, and if $Q\succsim P_n$ for all $n\in\N$, then $Q\succsim P$.
\end{axiom}

\begin{remark}
Suppose a branch with probability $p$ is split into two branches with probabilities $q$ and $r$, where $q+r=p$, and the prizes on the two branches converge to the same prize. The resulting split lotteries converge weakly to the coalesced lottery. Full continuity would therefore impose continuity across the splitting or coalescing of branches. We do not impose this requirement, since it would rule out discrete event-splitting effects, including those generated by support size costs.
\end{remark}

We impose continuity in probabilities, including when the probability of a prize approaches zero. This rules out discontinuous changes in valuation when a prize enters or leaves the support of a lottery. In \Cref{sec:extension:complex}, we relax this requirement to allow for support size costs, which may change discontinuously when the support changes.

\begin{axiom}[Mixture Continuity]
\label{ax:probcontinuity}
For all lotteries $P,Q$, and $R$, the sets
\[\big\{\alpha\in[0,1]\big| \alpha P+(1-\alpha)Q\succsim R\big\}\text{ and }\big\{\alpha\in[0,1]\big| R\succsim \alpha P+(1-\alpha)Q\big\}\text{ are closed.}\]
\end{axiom}

We impose strict first-order stochastic dominance only within a fixed induced probability tree. This captures the idea that comparing two lotteries is simpler when they share the same induced probability tree, since there are no differences in probability structure to evaluate. Hence, their ranking follows stochastic dominance.

\begin{axiom}[Within-Tree Strict FOSD]
\label{ax:samedistfosd}
If two lotteries $P$ and $Q$ have the same induced probability tree and $P$ strictly first-order stochastically dominates $Q$, then $P\succ Q$.
\end{axiom}

Branch cancellation is the key behavioral restriction across induced probability trees. It assumes that the decision maker cancels common prizes in comparisons. Hence, the ranking is invariant to the prize assigned to a common branch with a common probability. \Cref{fig:weakcindintro} illustrates the axiom.

\begin{axiom}[Branch Cancellation]
\label{ax:weakcindependence}
For all lotteries $P,Q$, prizes $x,y\notin\supp P\cup\supp Q$, and $\alpha\in(0,1)$,
\[\alpha P+(1-\alpha)\delta_x\succsim\alpha Q+(1-\alpha)\delta_x\iff\alpha P+(1-\alpha)\delta_y\succsim\alpha Q+(1-\alpha)\delta_y.\]
\end{axiom}

These axioms yield a perturbed expected utility representation for lotteries in which each branch has probability at most $1/2$. For branches with higher probabilities, however, probability and utility remain inseparable.

\begin{theorem}[Perturbed EU for Small Probabilities]{thm:variationalweightedfull}
$\succsim$ satisfies \Cref{ax:weakorder:simple,ax:weakcontinuityinmoney:simple,ax:probcontinuity,ax:weakcindependence,ax:samedistfosd} iff there exist $v:[0,1]\times X\to\R$ and a linearly continuous perturbation function $c:\Delta(X)\to\R$ such that:
\begin{enumerate}
\item for each $x\in X$, $v(\cdot,x)$ is continuous, and for each $p\in(0,1]$, $v(p,\cdot)$ is continuous and strictly increasing;
\item for each $p\leq1/2$ and $x\in X$, $v(p,x)=2p\,v(1/2,x)$;
\item for each $p>1/2$, the function $x\mapsto v(p,x)-v(1-p,x)$ is strictly increasing;
\end{enumerate}
and
\[P\succsim Q\iff\longremovespace\sum_{x\in\supp P}\longremovespace v(P(x),x)-c(P)\geq\longremovespace\sum_{x\in\supp Q}\longremovespace v(Q(x),x)-c(Q)\quad\text{for all }P,Q\in\Delta(X).\]
\end{theorem}

For $p\leq 1/2$, define $\tilde u(x)=2v(1/2,x)$. Then $v(p,x)=p\tilde u(x)$, so branches with probability at most $1/2$ enter in expected utility form. This scaling follows because branches with probabilities $p=r+s$, $r$, and $s$ can coexist in a common probability tree when $p<1/2$. Within-tree FOSD and continuity then imply that $v(p,x)=v(r,x)+v(s,x)$, yielding separability between probability and utility. For $p>1/2$, this argument no longer applies, and probability and utility remain potentially inseparable.

To extend expected utility scaling to these larger probabilities, we impose restricted common downscaling. This axiom requires that adding the same outside prize to two lotteries with the same induced probability tree cannot reverse their ranking. Since the lotteries share the same probability tree, their perturbation terms cancel from the comparison.

\begin{axiom}[Restricted Common Downscaling]
\label{ax:splittinghalf}
If two lotteries $P$ and $Q$ have the same induced probability tree, $x\notin\supp P\cup\supp Q$, and $\alpha\in(0,1)$, then $P\succsim Q$ implies $\alpha P+(1-\alpha)\delta_x\succsim\alpha Q+(1-\alpha)\delta_x$.
\end{axiom}

\begin{proposition}[Perturbed EU]{thm:variationalweighted}
$\succsim$ satisfies \Cref{ax:weakorder:simple,ax:weakcontinuityinmoney:simple,ax:probcontinuity,ax:weakcindependence,ax:samedistfosd,ax:splittinghalf} iff there exist a continuous and strictly increasing $u:X\to\R$ and a linearly continuous perturbation function $c:\Delta(X)\to\R$ such that
\[P\succsim Q\iff\longremovespace\sum_{x\in\supp P}\longremovespace P(x)u(x)-c(P)\geq\longremovespace\sum_{x\in\supp Q}\longremovespace Q(x)u(x)-c(Q)\quad\text{for all }P,Q\in\Delta(X).\]
\end{proposition}

\Cref{thm:variationalweightedfull} and \Cref{thm:variationalweighted} provide the behavioral foundations for perturbed expected utility. In this characterization, branch cancellation axiom separates the value of the induced probability tree from the value of the lottery. This is formalized in the next section. Whereas, within-tree FOSD and restricted common downscaling separates probabilities from utilities.\footnote{Uniqueness of the representation is established by \textcite[Theorem 5]{mononen2026preference}.}

This result provides a preference-based counterpart to \posscite{hofbauer2002} stochastic choice characterization for the perturbed expected utility under a known utility function. In contrast, our characterization derives both the utility function and a general perturbation function directly from preferences over lotteries.

\subsection{The Role of Within-Tree First-Order Stochastic Dominance}

We next examine what branch cancellation implies when within-tree FOSD is replaced by single prize monotonicity.
\setcounter{axiomp}{3}
\begin{axiomp}[Single Prize Monotonicity]
\label{ax:singleprizemonotonicity}
For all lotteries $P$, $\alpha\in(0,1]$, $x,y\notin\supp P$, $x>y$, we have
\[\alpha \delta_x +(1-\alpha) P\succ \alpha \delta_y +(1-\alpha) P.\]
\end{axiomp}

\begin{proposition}{prop:perturbedeuwithoutfosd}
$\succsim$ satisfies \Cref{ax:weakorder:simple,ax:weakcontinuityinmoney:simple,ax:probcontinuity,ax:weakcindependence,ax:singleprizemonotonicity} iff there exist $v:[0,1]\times X\to\R$ and a linearly continuous perturbation function $c:\Delta(X)\to\R$ such that:
\begin{enumerate}
\item for each $x\in X$, $v(\cdot,x)$ is continuous, and for each $p\in(0,1]$, $v(p,\cdot)$ is continuous and strictly increasing;
\end{enumerate}
and
\[P\succsim Q\iff\longremovespace\sum_{x\in\supp P}\longremovespace v(P(x),x)-c(P)\geq\longremovespace\sum_{x\in\supp Q}\longremovespace v(Q(x),x)-c(Q)\quad\text{for all }P,Q\in\Delta(X).\]
\end{proposition}

\Cref{prop:perturbedeuwithoutfosd} provides a representation associated with the cancellation of common outcome-probability pairs discussed by \textcite{kahneman1979}. Branch cancellation already separates the tree dependent perturbation from branch-specific prize evaluation. Without within-tree FOSD, however, probability and utility need not be separable in $v(p,x)$.  Within-tree FOSD imposes the additional restrictions on $v$ in \Cref{thm:variationalweightedfull} and, in particular, separates probability weights from prize utilities for branches with probabilities at most $1/2$.

\subsection{Additively Separable Perturbed Expected Utility}
The general perturbation allows interactions among branch probabilities: changing some branches can have a different structural effect depending on the remaining probability tree. We strengthen branch cancellation to lottery cancellation by requiring comparisons to be invariant to the choice of a common background lottery. This classical axiom from \textcite{machina1989dynamic} has substantial normative support \parencite{NielsenRehbeck2022}.
\setcounter{axiomp}{4}
\begin{axiomp}[Lottery Cancellation]
\label{ax:lotteryweakcindependence}
For all lotteries $P,Q,R,S$ with $(\supp P\cup \supp Q)\cap(\supp R\cup \supp S)=\emptyset$, and $\alpha\in(0,1)$,
\[\alpha P+(1-\alpha)R\succsim\alpha Q+(1-\alpha)R\iff\alpha P+(1-\alpha)S\succsim\alpha Q+(1-\alpha)S.\]
\end{axiomp}

\begin{proposition}[Additively Separable Perturbed EU]{thm:separablevariationalweighted}
$\succsim$ satisfies \Cref{ax:weakorder:simple,ax:weakcontinuityinmoney:simple,ax:probcontinuity,ax:lotteryweakcindependence,ax:samedistfosd,ax:splittinghalf} iff there exist a continuous and strictly increasing $u:X\to\R$ and a continuous function $\psi:[0,1]\to\R$ with $\psi(0)=0$ such that for all $P,Q\in\Delta(X)$,
\[P\succsim Q\iff\longremovespace\sum_{x\in\supp P}\longremovespace P(x)u(x)-\longremovespace\sum_{x\in\supp P}\longremovespace \psi\big(P(x)\big)+\psi(1)\geq\longremovespace\sum_{x\in\supp Q}\longremovespace Q(x)u(x)-\longremovespace\sum_{x\in\supp Q}\longremovespace \psi\big(Q(x)\big)+\psi(1).\]
\end{proposition}

\Cref{thm:separablevariationalweighted} shows that lottery cancellation rules out interactions among branch probabilities in the perturbation function. In particular, splitting a branch of probability $p$ into two branches with distinct prizes and probabilities $q$ and $p-q$ changes the perturbation by $\psi(q)+\psi(p-q)-\psi(p)$, independently of the remaining probability tree. Under the general representation in \Cref{thm:variationalweighted}, the effect of the same split can depend on the surrounding probability tree.

This result provides a preference-based counterpart to \posscite{fudenberg2015stochastic} stochastic choice characterization of additive perturbed utility. Additively separable perturbations, including entropy and quadratic specifications, are widely used in stochastic choice models, with applications to bounded rationality, learning in games, and discrete-choice modeling \parencite{strzalecki2025}.

\section{Support Size Representation}
\label{sec:extension:complex}

The general perturbation may depend on the entire distribution of branch probabilities. In this section, we characterize when the perturbation depends only on the number of distinct prizes, taking the support size form $c(P)=\kappa(|\supp P|)$. The number of distinct prizes is a natural measure of lottery complexity, used in models such as \textcite{puri2025}.

\begin{definition}
\label{def:supportsize-eu}
Preferences $\succsim$ admit a \emph{support size representation} if there exist a utility function $u:X\to\R$ and a function $\kappa:\N\to\R$ with $\kappa(1)=0$ such that, for all $P,Q\in\Delta(X),$
\[P\succsim Q\iff\longremovespace\sum_{x\in\supp P}\longremovespace P(x)u(x)-\kappa(|\supp P|)\geq\longremovespace\sum_{x\in\supp Q}\longremovespace Q(x)u(x)-\kappa(|\supp Q|),\]
\end{definition}

Support size perturbations can be discontinuous when a prize enters or leaves the support with vanishing probability. We therefore replace mixture continuity with unboundedness.
\begin{axiom}[Unboundedness]\label{ax:unbounded}
For all $\alpha\in[0,1)$, and lotteries $P,Q$, there exist $x,y\in X$ such that
\[\alpha P+(1-\alpha)\delta_x\succ Q\succ \alpha P+(1-\alpha)\delta_y.\]
\end{axiom}
This axiom guarantees that by changing a single prize in a lottery, we can make the lottery better or worse than any other fixed lottery.

We restrict the independence axiom to lotteries with the same support. When two lotteries have the same support, mixing each with a common lottery preserves the equality of their support sizes. Consequently, under a support size perturbation, the perturbation terms cancel from both the original comparison and the comparison between their mixtures. This motivates the following axiom.

\begin{axiom}[Same-Support Independence]\label{ax:samesupportind}
Assume that two lotteries $P,Q$ have the same support. Then for all lotteries $R$ and $\alpha\in(0,1)$
\[P\succsim Q \iff \alpha P+(1-\alpha)R\succsim \alpha Q+(1-\alpha)R\]
\end{axiom}

\begin{theorem}[Support Size Representation and Certainty Effect]{thm:costforsupport}
$\succsim$ satisfies \Cref{ax:weakorder:simple,ax:weakcontinuityinmoney:simple,ax:unbounded,ax:weakcindependence,ax:samedistfosd,ax:samesupportind} iff there exist an unbounded, continuous, and strictly increasing function $u:X\to\R$, a continuous and strictly increasing function $\phi:\R\to\R$, and a function $\kappa:\N\to\R$ such that the function $V:\Delta(X)\to\R$ defined by \[V(P)=\begin{dcases*} \sum_{x\in\supp P}P(x)u(x)-\kappa(|\supp P|), &$|\supp P|>1$\\ \phi(u(x)), &\text{if }$P=\delta_x$ \end{dcases*}\] represents $\succsim$:
\[P\succsim Q\iff V(P)\geq V(Q)\quad\text{ for all $P,Q\in\Delta(X)$.}\]
\end{theorem}

Although same-support independence applies only to comparisons between lotteries with identical supports, \Cref{thm:costforsupport} shows that it restricts the perturbation to depend only on support size across all nondegenerate lotteries. However, certainties may be evaluated using a different utility function than risky lotteries. The transformation $\phi$ can be nonlinear, allowing the model to accommodate a certainty effect that varies with the prize level in addition to the support size perturbation. A related model of the certainty effect, without a support size perturbation, has been studied by \textcite{SCHMIDT199832}.

To link the valuation of certainty to that of risky lotteries, we impose the following consistency condition on certainty equivalents.

\begin{axiom}[Certainty Equivalent Swap]\label{ax:certaintyequivalentswap}
Assume that lotteries $P$ and $Q$ have the same support and $x,y\notin\supp P\cup \supp Q$. If $P\sim \delta_x$ and $Q\sim \delta_y$, then
\[\frac12 P+\frac12 \delta_y\sim \frac12 Q+\frac12 \delta_x.\]
\end{axiom}

\begin{proposition}[Support Size Representation]{prop:costforsupportfull}
$\succsim$ satisfies \Cref{ax:weakorder:simple,ax:weakcontinuityinmoney:simple,ax:unbounded,ax:weakcindependence,ax:samedistfosd,ax:samesupportind,ax:certaintyequivalentswap} iff there exist an unbounded, continuous, and strictly increasing function $u:X\to\R$ and  a function $\kappa:\N\to\R$ with $\kappa(1)=0$ such that, for all $P,Q\in\Delta(X)$,
\[P\succsim Q\iff\!\!\! \sum_{x\in\supp P}\!\!P(x)u(x)-\kappa(|\supp P|)\geq \!\!\sum_{x\in\supp Q}\!\!Q(x)u(x)-\kappa(|\supp Q|).\]
\end{proposition}

\Cref{prop:costforsupportfull} shows that certainty equivalent swap eliminates the separate evaluation of certainties allowed by \Cref{thm:costforsupport}. Mixing each of two lotteries with the other's certainty equivalent yields lotteries with the same support size, so their perturbation terms cancel. The required indifference then equates the utility difference between the certainty equivalents with the expected utility difference between the original lotteries. This forces the transformation $\phi$ to differ from the identity only by an additive constant, which can be absorbed into the support size perturbation.

\textbf{Comparison to \textcite{puri2025}.} Puri characterizes weakly increasing support size costs using same-support independence and a differences-in-differences complexity aversion axiom. The latter compares two lotteries after mixing each with the other's certainty equivalent and then with a common lottery whose support contains all relevant prizes. Exchanging certainty equivalents offsets the differences in the original valuations, while the common lottery equalizes the support sizes of the resulting lotteries. The comparison therefore isolates the effect of the difference in support size between the original lotteries.

Our characterization provides an alternative axiomatic foundation that decomposes the support size representation. Branch cancellation yields a general perturbed utility representation with nonseparable utility. Adding within-tree FOSD separates probabilities and utilities. Adding same-support independence reduces the perturbation to a support size term, but certainty and risk may still be evaluated differently. Certainty equivalent swap aligns the evaluation of certainty and risk. Finally, adding concavity characterizes weakly increasing support size costs, as shown in \Cref{sec:preference_for_simplicity}. This decomposition connects the support size representation to the general perturbed expected utility representation and derives its restrictions from familiar behavioral principles in choice under risk.

The characterizations also differ in their regularity assumptions. In particular, our unboundedness axiom guarantees the existence of certainty equivalents for all lotteries. We provide a counterexample to Puri's Theorem 1 in \Cref{app:puricounterexample}, showing that her stated axioms permit preferences that do not admit a support size representation.

\section{Preference for Simplicity}
\label{sec:preference_for_simplicity}

The preceding characterizations identify which features of the induced probability tree can affect value, while leaving the direction of that dependence unrestricted. We now require that mixing a preferred lottery with a weakly worse lottery cannot produce a lottery preferred to the better alternative.

\begin{axiom}[Concavity]\label{ax:convexity}
For all $P,Q\in\Delta(X)$ and $\alpha\in(0,1)$, if $P\succsim Q$, then
\[P\succsim \alpha P+(1-\alpha)Q.\]
\end{axiom}

\begin{proposition}[Concavity of the Perturbation Function]{thm:concave-information}
Assume that $\succsim$ has a perturbed expected utility representation by a continuous and strictly increasing $u:X\to\R$ such that $u(X)=\R$, and a perturbation function $c:\Delta(X)\to\R$. Then $\succsim$ satisfies \Cref{ax:convexity} if and only if $c$ is concave.
\end{proposition}

Since $c$ vanishes at certainty, concavity makes it nonnegative and weakly increasing under branch splitting. Concavity also restricts costs without changing support size: permuting two branch probabilities leaves $c$ unchanged, while averaging these permutations equalizes the probabilities and weakly increases $c$. Averaging over all permutations therefore shows that the uniform probability tree has weakly maximal cost among trees with the same number of branches. At equal expected utility, a lottery with uniform branch probabilities is consequently weakly less preferred than one with nonuniform probabilities and the same number of prizes.

For the support size representation, the same concavity axiom characterizes when the perturbation is weakly increasing in the number of distinct prizes.

\begin{corollary}[Monotone Support Size Costs]{cor:monotone-support-size-costs}
Suppose that $\succsim$ has a support size representation by a continuous and strictly increasing $u:X\to\R$, and $\kappa:\N\to\R$. Then $\succsim$ satisfies \Cref{ax:convexity} iff $\kappa$ is weakly increasing.
\end{corollary}

Unlike the general perturbation, the support size cost is unaffected by redistributions of probability among existing prizes. Concavity instead requires that introducing a distinct possible prize weakly increases the cost, even when its probability is arbitrarily small.

\section{Examples and Applications}
\label{sec:examples_applications}

The characterizations accommodate both separable and nonseparable perturbations. In the separable class, $c(P)=\sum_{x\in\supp P}\psi(P(x))-\psi(1)$ for a continuous $\psi$ with $\psi(0)=0$. For $\gamma>0$, examples include Shannon entropy $c(P)=\gamma H(P)$, where $H(P)=-\sum_{x\in\supp P}P(x)\log P(x)$, and the quadratic cost $c(P)=\gamma(1-\sum_{x\in\supp P}P(x)^2)$.

The general representation also permits nonseparable perturbations. Examples include square-root Shannon entropy, $c(P)=\gamma\sqrt{H(P)}$ for $\gamma>0$; norm entropies, $c(P)=1-\big(\sum_{x\in\supp P}P(x)^q\big)^{1/q}$ for $q>1$; and Rényi entropies, $c(P)=\frac{1}{1-\beta}\log\big(\sum_{x\in\supp P}P(x)^\beta\big)$ for $\beta>0$, $\beta\neq1$ \parencite{blondel2020learning,bilodeau2023rootentropy}. Unlike additively separable perturbations, the effect of splitting a branch under these specifications can depend on the probabilities of the remaining branches.

Support size perturbations impose a different restriction: $c(P)=\kappa(|\supp P|)$ depends only on the number of distinct prizes. Examples include $\kappa(n)=\gamma\log n$ and $\kappa(n)=\gamma\mathbf{1}\{n\geq2\}$ for $\gamma\geq0$. A nonzero support size increment persists even when the probability of a newly introduced prize approaches zero, unlike the preceding linearly continuous perturbations.

\subsection{Event Splitting and Coalescence}
\label{sec:event-splitting}

The split event experiment of \textcite{bernheim2020} provides an application of the support size representation. In their split-low treatment, the baseline lottery is $P_0=.4\delta_{30}+.6\delta_{20}$, while the split lottery is $P_\epsilon=.4\delta_{30}+.3\delta_{20-\epsilon}+.3\delta_{20+\epsilon}$, with $\epsilon\in\{0.5,1,2,3\}$.
The reduced lottery has support size two at $\epsilon=0$ and support size three for every $\epsilon>0$. Bernheim and Sprenger also present the baseline as a $30$-$30$-$40$ lottery with identical prizes on the two $30\%$ events and obtain essentially the same certainty equivalents as under the $60$-$40$ presentation. Their design therefore separates the creation of an additional distinct prize from the presentation of an additional event.

Let $x_\epsilon$ be the certainty equivalent of $P_\epsilon$. Under \Cref{prop:costforsupportfull}, for every $\epsilon>0$,
\[u(x_0)-u(x_\epsilon)=\kappa(3)-\kappa(2)+.3\bigl[2u(20)-u(20-\epsilon)-u(20+\epsilon)\bigr].\]
The first term is the increase in the support size cost, while the second is the expected utility effect of separating the two prizes. As $\epsilon\downarrow0$, the latter vanishes, so
\[u(x_0)-\lim_{\epsilon\downarrow0}u(x_\epsilon)=\kappa(3)-\kappa(2).\]
Thus, a positive support size increment generates a discontinuous decline in the certainty equivalent when the two prizes become distinct. For every positive $\epsilon$, support size remains three, so further variation in the certainty equivalent comes only through expected utility.

\textcite{bernheim2020} report that increasing $\epsilon$ from $0$ to $0.5$ reduces the average certainty equivalent by $\$0.47$, with a standard error of $0.11$, after which certainty equivalents level off as $\epsilon$ increases. They find a qualitatively similar, though smaller, effect when splitting the high prize. This pattern is consistent with a positive support size increment $\kappa(3)-\kappa(2)$, as implied by \Cref{cor:monotone-support-size-costs}. However, the observed change at $\epsilon=0.5$ also includes the expected utility effect of separating the prizes.\footnote{\textcite{bernheim2020} estimate power utility $u(x)=x^\alpha$ with $\alpha=0.975$ in this treatment. Using this estimate as a local calibration, the expected-utility effect of replacing the $60\%$ chance of $20$ by $30\%$ at $19.5$ and $30\%$ at $20.5$ changes the certainty equivalent by less than one cent, compared with the observed $\$0.47$ decline.}

\subsection{Deductible insurance}
\label{sec:deductible-insurance}

Deductible insurance provides an application in which changes in the deductible can alter the induced probability tree of final wealth by merging distinct loss states into the same outcome.\footnote{Deductible contracts are a classical object in insurance theory \parencite{arrow1974}. Deductible choices are also widely used to study risk aversion and probability weighting \parencite{cohen2007,sydnor2010,barseghyan2013}. Related work studies complexity in insurance-plan selection \parencite{besliu2022} and consumers' difficulty mapping insurance contracts into distributions of final financial consequences \parencite{samek2025}.}
Suppose that losses $0=\ell_1<\cdots<\ell_n$ occur with probabilities $p_1,\ldots,p_n>0$, and initial wealth is $m$. A contract with deductible $d$ requires the insured to bear the first $d$ units of a loss and covers any loss above $d$. If its premium is $\pi(d)$, final wealth in loss state $\ell_i$ is $m-\pi(d)-\min\{\ell_i,d\}$.

If $\ell_k<d<\ell_{k+1}$, let $q_k=\sum_{i=k+1}^n p_i$. The reduced final wealth lottery is
\[P_d=\sum_{i=1}^k p_i\delta_{m-\pi(d)-\ell_i}+q_k\delta_{m-\pi(d)-d}.\]
All losses exceeding $d$ generate the same final wealth, so their branches coalesce in the reduced lottery, whose induced probability tree is $(p_1,\ldots,p_k,q_k)$.

Under perturbed utility, deductible insurance is evaluated by $V(P_d)=\sum_{x\in\supp P_d}P_d(x)u(x)-c(P_d)$. As long as $d\in(\ell_k,\ell_{k+1})$, the induced probability tree remains unchanged, so $c(P_d)$ is constant. When the deductible falls to $\ell_k$, the branches with probabilities $p_k$ and $q_k$ coalesce into a single branch with probability $p_k+q_k$. If $\pi$ is continuous, continuity of $u$ implies
\[V(P_{\ell_k})-\lim_{d\searrow\ell_k}V(P_d)=\lim_{d\searrow\ell_k}c(P_d)-c(P_{\ell_k}).\]
Thus, any discontinuity in valuation at the threshold comes entirely from the perturbation. Under concavity, merging the branches weakly reduces $c$, generating a weakly positive jump in value.

Different perturbation specifications yield different magnitudes of this gain. For the Shannon-entropy cost $c(P)=\gamma H(P)$, $\gamma>0$, coalescing the branches $p_k$ and $q_k$ reduces the cost by $\gamma[-p_k\log p_k-q_k\log q_k+(p_k+q_k)\log(p_k+q_k)]$. For the quadratic cost $c(P)=\gamma(1-\sum_xP(x)^2)$, $\gamma>0$, the corresponding reduction is $2\gamma p_kq_k$.

Under \Cref{prop:costforsupportfull}, the support size cost falls from $\kappa(k+1)$ to $\kappa(k)$ when the deductible reaches $\ell_k$. The resulting gain, $\kappa(k+1)-\kappa(k)$, is nonnegative under \Cref{cor:monotone-support-size-costs}. Unlike the entropy and quadratic specifications, the support size gain is independent of the probabilities of the merged branches. Thus, the different specifications yield distinct predictions about how the value of coalescence varies with the distribution of losses.

\section{Conclusion}\label{sec:conclusion:complex}

This paper provides behavioral foundations for separating the valuation of final prizes from the effects of a lottery's induced probability tree. The characterizations connect different forms of tree dependence to familiar principles of cancellation, dominance, and independence. General perturbations allow interactions among branch probabilities, additive separability eliminates these interactions, and support size dependence makes the perturbation independent of branch probabilities. An additional concavity axiom characterizes when these perturbations represent a preference for simplicity. Our applications to event splitting and deductible insurance illustrate how different complexity specifications can generate distinct predictions.

{\sloppy\printbibliography}

@article{lucegamble12008,
 author = {R. Duncan Luce and C. T. Ng and A. A. J. Marley and János Aczél},
 journal = {Economic Theory},
 number = {1},
 pages = {1--33},
 publisher = {Springer},
 title = {Utility of gambling I: entropy modified linear weighted utility},
 volume = {36},
 year = {2008}
}

@article{lucegamble22008,
 author = {R. Duncan Luce and C. T. Ng and A. A. J. Marley and János Aczél},
 journal = {Economic Theory},
 number = {2},
 pages = {165--187},
 publisher = {Springer},
 title = {Utility of gambling II: risk, paradoxes, and data},
 volume = {36},
 year = {2008}
}

@article{levin2016,
  title={Ellsberg paradox: Ambiguity and complexity aversions compared},
%  author={Kov{\'a}{\v{r}}{\'\i}k, Jarom{\'\i}r and Levin, Dan and Wang, Tao},
  author={Kovářík, Jaromír and Levin, Dan and Wang, Tao}, 
  journal={Journal of Risk and Uncertainty},
  volume={52},
  number={1},
  pages={47--64},
  year={2016}
}

@article{huck1999,
title = "Risk, complexity, and deviations from expected-value maximization: Results of a lottery choice experiment",
journal = "Journal of Economic Psychology",
volume = "20",
number = "6",
pages = "699 - 715",
year = "1999",
author = "Steffen Huck and Georg Weizsäcker"
}

@article{bernheim2020,
author = {Bernheim, B. Douglas and Sprenger, Charles},
title = {On the Empirical Validity of Cumulative Prospect Theory: Experimental Evidence of Rank-Independent Probability Weighting},
journal = {Econometrica},
volume = {88},
number = {4},
pages = {1363-1409},
year = {2020}
}

@article{stodder1997,
 ISSN = {00945056, 19394632},
 URL = {http://www.jstor.org/stable/40325750},
 author = {James Stodder},
 journal = {Eastern Economic Journal},
 number = {1},
 pages = {1--15},
 publisher = {Palgrave Macmillan Journals},
 title = {Complexity Aversion: Simplification in the Herrnstein and Allais Behaviors},
 volume = {23},
 year = {1997}
}

@book{krantz1971,
  title={Foundations of Measurement. Vol. I. Additive and Polynomial Representations},
  author={Krantz, David and Luce, Duncan and Suppes, Patrick and Tversky, Amos},
  year={1971},
 publisher = {Academic Press}
}

@article{kahneman1979,
 author = {Daniel Kahneman and Amos Tversky},
 journal = {Econometrica},
 number = {2},
 pages = {263--291},
 title = {Prospect Theory: An Analysis of Decision under Risk},
 volume = {47},
 year = {1979}
}

@article{birnbaum2008,
  author={Michael H. Birnbaum},
 journal = {Psychological Review},
 number = {2},
 pages = {463--501},
  title = {New Paradoxes of Risky Decision Making},
 volume = {115},
 year = {2008}
}

@article{mador2000,
title = {On complexity and lotteries' evaluation – three experimental observations},
journal = {Journal of Economic Psychology},
volume = {21},
number = {6},
pages = {625-637},
year = {2000},
author = {Galit Mador and Doron Sonsino and Uri Benzion}
}

@article{moffatt2015,
title = {Heterogeneity in preferences towards complexitys},
journal = {Journal of Risk and Uncertainty},
volume = {51},
number = {2},
pages = {147--170},
year = {2015},
author = {Moffatt, Peter G. and Sitzia, Stefania and Zizzo, Daniel John}
}

@article{maccheroni2006ambiguity,
  title={Ambiguity Aversion, Robustness, and the Variational Representation of Preferences},
  author={Maccheroni, Fabio and Marinacci, Massimo and Rustichini, Aldo},
  journal={Econometrica},
  volume={74},
  number={6},
  pages={1447--1498},
  year={2006}
}

@article{diecidue2004,
 author = {Enrico Diecidue and Ulrich Schmidt and Peter P. Wakker},
 journal = {Journal of Risk and Uncertainty},
 number = {3},
 pages = {241--259},
 title = {The Utility of Gambling Reconsidered},
 volume = {29},
 year = {2004}
}

@article{menestrel2001,
 author = {Le Menestrel, Marc},
 journal = {Theory and Decision},
 number = {3},
 pages = {249--262},
 title = {A Process Approach to the Utility for Gambling},
 volume = {50},
 year = {2001}
}

@article{conlisk1993,
 author = {John Conlisk},
 journal = {Journal of Risk and Uncertainty},
 number = {3},
 pages = {255--275},
 publisher = {Springer},
 title = {The Utility of Gambling},
 volume = {6},
 year = {1993}
}

@article{mononenprospecttheory,
  author  = {Mononen, Lasse},
  title   = {An Axiomatization of Separable Prospect Theory},
  journal = {Journal of Risk and Uncertainty},
  volume  = {},
  pages   = {251--275},
  year    = {2026},
  doi     = {10.1007/s11166-026-09494-w}
}

@article{fudenberg2015stochastic,
  title={Stochastic choice and revealed perturbed utility},
  author={Fudenberg, Drew and Iijima, Ryota and Strzalecki, Tomasz},
  journal={Econometrica},
  volume={83},
  number={6},
  pages={2371--2409},
  year={2015},
  publisher={Wiley Online Library}
}

@article{handa1977,
  title={Risk, probabilities, and a new theory of cardinal utility},
  author={Handa, Jagdish},
  journal={Journal of Political Economy},
  volume={85},
  number={1},
  pages={97--122},
  year={1977},
  publisher={The University of Chicago Press}
}

@article{mononen2026preference,
  title={On preference for simplicity and probability weighting},
  author={Mononen, Lasse},
  journal={Journal of Economic Theory},
  pages={106225},
  year={2026},
  publisher={Elsevier}
}

@article{puri2025,
  title={Simplicity and risk},
  author={Puri, Indira},
  journal={The Journal of Finance},
  volume={80},
  number={2},
  pages={1029--1080},
  year={2025},
  publisher={Wiley Online Library}
}

@article{hofbauer2002,
  title={On the global convergence of stochastic fictitious play},
  author={Hofbauer, Josef and Sandholm, William H},
  journal={Econometrica},
  volume={70},
  number={6},
  pages={2265--2294},
  year={2002},
  publisher={Wiley Online Library}
}

@article{SCHMIDT199832,
title = {A Measurement of the Certainty Effect},
journal = {Journal of Mathematical Psychology},
volume = {42},
number = {1},
pages = {32-47},
year = {1998},
author = {Ulrich Schmidt}
}

@article{NielsenRehbeck2022,
  author  = {Nielsen, Kirby and Rehbeck, John},
  title   = {When Choices Are Mistakes},
  journal = {American Economic Review},
  year    = {2022},
  volume  = {112},
  number  = {7},
  pages   = {2237--2268},
  doi     = {10.1257/aer.20201550}
}

@article{arrow1974,
  title={Optimal insurance and generalized deductibles},
  author={Arrow, Kenneth J},
  journal={Scandinavian Actuarial Journal},
  volume={1974},
  number={1},
  pages={1--42},
  year={1974},
  publisher={Taylor \& Francis}
}

@article{cohen2007,
  title={Estimating risk preferences from deductible choice},
  author={Cohen, Alma and Einav, Liran},
  journal={American economic review},
  volume={97},
  number={3},
  pages={745--788},
  year={2007},
  publisher={American Economic Association}
}

@article{sydnor2010,
  title={(Over) insuring modest risks},
  author={Sydnor, Justin},
  journal={American Economic Journal: Applied Economics},
  volume={2},
  number={4},
  pages={177--199},
  year={2010},
  publisher={American Economic Association}
}

@article{barseghyan2013,
  title={The nature of risk preferences: Evidence from insurance choices},
  author={Barseghyan, Levon and Molinari, Francesca and O'Donoghue, Ted and Teitelbaum, Joshua C},
  journal={American economic review},
  volume={103},
  number={6},
  pages={2499--2529},
  year={2013},
  publisher={American Economic Association}
}

@article{besliu2022,
  title={Complexity in insurance selection: Cross-classified multilevel analysis of experimental data},
  author={Be{\c{s}}liu, Corina},
  journal={Journal of Behavioral and Experimental Finance},
  volume={35},
  pages={100713},
  year={2022},
  publisher={Elsevier}
}

@article{samek2025,
  title={The impact of consequence information on insurance choice},
  author={Samek, Anya and Sydnor, Justin},
  journal={Review of Economics and Statistics},
  pages={1--46},
  year={2025},
  publisher={MIT Press, 255 Main Street, 9th Floor, Cambridge, Massachusetts 02142, USA~…}
}

@article{oberholzer2024complexity,
  title={Complexity aversion in risky choices and valuations: Moderators and possible causes},
  author={Oberholzer, Yvonne and Olschewski, Sebastian and Scheibehenne, Benjamin},
  journal={Journal of Economic Psychology},
  volume={100},
  pages={102681},
  year={2024},
  publisher={Elsevier}
}

@unpublished{principi2026distance,
  title={A distance-based theory of lottery complexity},
  author={Principi, Giulio},
  journal={arXiv preprint arXiv:2608.14464},
  year={2026}
}

@article{goodmanpuri2025,
  title={Overvaluing simple bets: Evidence from the options market},
  author={Goodman, Aaron and Puri, Indira},
  journal={Journal of Financial Economics},
  volume={172},
  pages={104140},
  year={2025},
  publisher={Elsevier}
}

@article{deBruijn1951,
   author  = {{de Bruijn}, N. G.},
   title   = {Functions whose differences belong to a given class},
   journal = {Nieuw Archief voor Wiskunde},
   series  = {2},
   volume  = {23},
   year    = {1951},
   pages   = {194--218}}

@article{ng1974,
  author  = {Ng, C. T.},
  title   = {Representation for Measures of Information with
             the Branching Property},
  journal = {Information and Control},
  volume  = {25},
  number  = {1},
  pages   = {45--56},
  year    = {1974},
  doi     = {10.1016/S0019-9958(74)90788-8}
}

@article{ebanks1981,
  author  = {Ebanks, Bruce},
  title   = {A Characterization of Separable Utility Functions},
  journal = {Kybernetika},
  volume  = {17},
  number  = {3},
  pages   = {244--255},
  year    = {1981}
}

@article{machina1989dynamic,
  author  = {Machina, Mark J.},
  title   = {Dynamic Consistency and Non-Expected Utility Models of Choice under Uncertainty},
  journal = {Journal of Economic Literature},
  year    = {1989},
  volume  = {27},
  number  = {4},
  pages   = {1622--1668}
}

@book{strzalecki2025,
  author    = {Strzalecki, Tomasz},
  title     = {Stochastic Choice Theory},
  year      = {2025},
  publisher = {Cambridge University Press},
  series    = {Econometric Society Monographs},
  doi       = {10.1017/9781009512749},
  isbn      = {9781009512763}
}

@article{bilodeau2023rootentropy,
  author  = {Bilodeau, Blair and Negrea, Jeffrey
             and Roy, Daniel M.},
  title   = {Relaxing the i.i.d. Assumption:
             Adaptively Minimax Optimal Regret
             via Root-Entropic Regularization},
  journal = {The Annals of Statistics},
  year    = {2023},
  volume  = {51},
  number  = {4},
  pages   = {1850--1876},
  doi     = {10.1214/23-AOS2315}
}

@article{blondel2020learning,
  author  = {Blondel, Mathieu and Martins, Andr{\'e} F. T. and Niculae, Vlad},
  title   = {Learning with {Fenchel--Young} Losses},
  journal = {Journal of Machine Learning Research},
  volume  = {21},
  number  = {35},
  pages   = {1--69},
  year    = {2020},
  url     = {https://jmlr.org/papers/v21/19-021.html}
}

@article{blavatskyy2008stochastic,
  author  = {Blavatskyy, Pavlo R.},
  title   = {Stochastic Utility Theorem},
  journal = {Journal of Mathematical Economics},
  year    = {2008},
  volume  = {44},
  number  = {11},
  pages   = {1049--1056},
  doi     = {10.1016/j.jmateco.2007.12.005}
}

@article{ke2018rational,
  author  = {Ke, Shaowei},
  title   = {Rational Expectation of Mistakes and a Measure of Error-Proneness},
  journal = {Theoretical Economics},
  year    = {2018},
  volume  = {13},
  number  = {2},
  pages   = {527--552},
  doi     = {10.3982/TE2477}
}
\newpage

\appendix
\numberwithin{equation}{section} 

\titlespacing*{\section}{0pt}{0.9ex plus 0.5ex minus 0.7ex}{0.9ex plus 0.5ex minus 0.7ex}
\titlespacing*{\subsection}{20pt}{0.9ex plus 0.5ex minus 0.7ex}{0.9ex plus 0.5ex minus 0.7ex}
\titlespacing*{\subsubsection}{30pt}{0.9ex plus 0.5ex minus 0.7ex}{0.9ex plus 0.5ex minus 0.7ex}
\clubpenalty=1
\widowpenalty=1
\displaywidowpenalty=1
\sloppy

{\normalfont\center\Large\bfseries Appendix to ``Decomposing the Cost of Complexity Under Risk'' \par}\vspace{6pt}

We represent induced probability trees as finite-support distributions on indices:
\[\Delta(\N)=\left\{p\in\R_+^{\N}:\sum_{i\in\N}p_i=1,\ |\supp p|<\infty\right\}.\]
Let $\Delta^4(\N)=\{p\in\Delta(\N):|\supp p|\geq4\}$ and, for $p\in\Delta(\N)$, let
$\ndiag{p}=\{x\in X^{\supp p}:x_i\neq x_j\text{ for all distinct }i,j\in\supp p\}$.
For $p\in\Delta(\N)$ and $x\in\ndiag{p}$, write $(p,x)=\sum_{i\in\supp p}p_i\delta_{x_i}$.
For each $p\in\Delta(\N)$, define $\succsim_p$ on $\ndiag{p}$ by
$x\succsim_p y$ if and only if $(p,x)\succsim(p,y)$.

\section{Proof of \Cref{thm:variationalweightedfull} and \Cref{thm:variationalweighted}}

The following result follows from \textcite{mononenprospecttheory}.

\begin{lemma}{lem:smooth-large-support-tree-cost}
If $\succsim$ satisfies \Cref{ax:weakorder:simple,ax:weakcontinuityinmoney:simple,ax:probcontinuity,ax:weakcindependence,ax:samedistfosd}, then there exist
$x^0\in\interior X$, continuous and strictly increasing functions
$v(a,\cdot):X\to\R$ for each $a\in(0,1)$ with $v(a,x^0)=0$, and a
perturbation function $\bar c:\Delta(X)\to\R$ such that, for all simple
lotteries $P,Q$ with $|\supp P|,|\supp Q|\geq4$,
\[P\succsim Q \iff \sum_{x\in\supp P}v(P(x),x)-\bar c(P) \geq \sum_{x\in\supp Q}v(Q(x),x)-\bar c(Q).\]
\end{lemma}

\begin{lemma}{lem:smooth-small-prob-linearity}
The representation in \Cref{lem:smooth-large-support-tree-cost} can be chosen so that $v(p,x)=4p\,v(1/4,x)$ for all $p\in(0,1/2)$ and $x\in X$.
\end{lemma}

\begin{proof}
Fix $r,s>0$ with $r+s=p<1/2$. We first show that $v(p,x)=v(r,x)+v(s,x)$ for every $x\in X$. The claim is immediate at $x=x^0$, so fix $x\neq x^0$ and let $q=(p,r,s,1-2p)\in\Delta^4(\N)$. Since $X$ is open, choose positive sequences $\varepsilon_k^1>\varepsilon_k^2>\varepsilon_k^3\to0$ such that
\[z_k^+=(x^0+\varepsilon_k^1,x+\varepsilon_k^2,x+\varepsilon_k^3,x^0),\quad z_k=(x,x^0+\varepsilon_k^2,x^0+\varepsilon_k^3,x^0),\quad z_k^-=(x^0-\varepsilon_k^1,x-\varepsilon_k^2,x-\varepsilon_k^3,x^0)\]
belong to $\ndiag{q}$.

Since $p=r+s$, $(q,z_k^+)$ strictly first-order stochastically dominates $(q,z_k)$, which strictly first-order stochastically dominates $(q,z_k^-)$. By \Cref{ax:samedistfosd} and cancellation of the common perturbation term,
\[\begin{aligned}v(p,x^0+\varepsilon_k^1)+v(r,x+\varepsilon_k^2)+v(s,x+\varepsilon_k^3) &>v(p,x)+v(r,x^0+\varepsilon_k^2)+v(s,x^0+\varepsilon_k^3)\\ &>v(p,x^0-\varepsilon_k^1)+v(r,x-\varepsilon_k^2)+v(s,x-\varepsilon_k^3).\end{aligned}\]
Letting $k\to\infty$ and using continuity and $v(a,x^0)=0$ gives $v(p,x)=v(r,x)+v(s,x)$.

Thus, for each fixed $x\neq x^0$, either $p\mapsto v(p,x)$ or $p\mapsto -v(p,x)$ is positive and additive on $(0,1/2)$. It is therefore increasing and, by the standard monotone Cauchy argument, linear in $p$. Normalizing at $p=1/4$ yields $v(p,x)=4p\,v(1/4,x)$. The same equality holds at $x=x^0$ by normalization.
\end{proof}

\begin{lemma}{lem:smooth-prob-continuity}
Under the assumptions and notation of \Cref{lem:smooth-small-prob-linearity}, $v(\cdot,x)$ is continuous on $(0,1)$ for every $x\in X$. Moreover, setting $v(0,x)=0$ gives a continuous extension to $[0,1)$.
\end{lemma}

\begin{proof}
By \Cref{lem:smooth-small-prob-linearity}, continuity holds on $(0,1/2)$. Fix $p\in[1/2,1)$, and choose $\epsilon>0$ such that, for $t$ sufficiently close to $p$, $q(t)=(t,1-t-2\epsilon,\epsilon,\epsilon)\in\Delta^4(\N)$ and $1-t-2\epsilon<1/2$. Write $\bar c(q(t))$ for the perturbation term associated with this tree.

We first show that $\bar c(q(t))\to \bar c(q(p))$. Choose $x=(x^0,y,x_3,x_4)\in\ndiag{q(p)}$. By \Cref{lem:smooth-small-prob-linearity}, the branch-utility sum of $(q(t),x)$ is
\[A(t)=4(1-t-2\epsilon)v(1/4,y) +4\epsilon v(1/4,x_3)+4\epsilon v(1/4,x_4),\]
which is continuous in $t$.

Suppose instead that, along some sequence $t_n\to p$, $\bar c(q(t_n))\geq \bar c(q(p))+\delta$ for some $\delta>0$. For large $n$, $|A(t_n)-A(p)|<\delta/4$, so the represented value of $(q(t_n),x)$ is at most $A(p)-\bar c(q(p))-3\delta/4$. Choose $z^-\in\ndiag{q(p)}$ with $z_i^-<x_i$ for every $i$ and
\[0<\sum_{i=1}^4\bigl(v(q_i(p),x_i)-v(q_i(p),z_i^-)\bigr)<\delta/4.\]
Then $(q(p),x)\succ(q(p),z^-)\succ(q(t_n),x)$ for all large $n$, contradicting \Cref{ax:probcontinuity}. The case $\bar c(q(t_n))\leq \bar c(q(p))-\delta$ is symmetric, using a sufficiently small coordinatewise upward perturbation of $x$. Hence $\bar c(q(t))\to \bar c(q(p))$.

We next show that $v(t,a)\to v(p,a)$ for every $a\in X$. This is immediate for $a=x^0$. Fix $a\neq x^0$ and choose $x_3,x_4$ such that $\hat x=(a,x^0,x_3,x_4)\in\ndiag{q(p)}$. The represented value of $(q(t),\hat x)$ is
\[v(t,a)+4\epsilon v(1/4,x_3)+4\epsilon v(1/4,x_4)-\bar c(q(t)).\]
Suppose, for example, that $v(t_n,a)\geq v(p,a)+\delta$ along some $t_n\to p$. Choose $b>a$ sufficiently close to $a$ that $v(p,b)-v(p,a)<\delta/4$ and $(b,x^0,x_3,x_4)\in\ndiag{q(p)}$. Since $\bar c(q(t_n))\to \bar c(q(p))$, for all large $n$,
\[(q(t_n),\hat x)\succ(q(p),(b,x^0,x_3,x_4)) \succ(q(p),\hat x),\]
contradicting \Cref{ax:probcontinuity}. The case $v(t_n,a)\leq v(p,a)-\delta$ is symmetric. Thus $v(t,a)\to v(p,a)$.

Hence $v(\cdot,x)$ is continuous on $(0,1)$. Continuity at $1/2$, together with \Cref{lem:smooth-small-prob-linearity}, gives $v(1/2,x)=2v(1/4,x)$, so $v(p,x)=4p\,v(1/4,x)$ for all $p\in[0,1/2]$ after setting $v(0,x)=0$. Continuity at $0$ follows.
\end{proof}

\begin{lemma}{lem:smooth-extension}
Under the assumptions and notation of \Cref{lem:smooth-prob-continuity}, $v(\cdot,x)$ has a continuous extension to $[0,1]$ for every $x\in X$, and there exists a linearly continuous perturbation function $c:\Delta(X)\to\R$ such that, for all $P,Q\in\Delta(X)$,
\[P\succsim Q\iff\sum_{x\in\supp P}v(P(x),x)-c(P)\geq\sum_{x\in\supp Q}v(Q(x),x)-c(Q).\]
\end{lemma}

\begin{proof}
Let $P=(p,x)$. Choose a four-prize lottery $R$ with $\supp R\cap\supp P=\emptyset$ and $\eta>0$ such that $x\pm\eta\mathbf 1\in\ndiag{p}$ and all intermediate prizes remain outside $\supp R$. Setting $P^\pm=(p,x\pm\eta\mathbf 1)$, \Cref{ax:samedistfosd} gives $P^+\succ P\succ P^-$. Hence, by \Cref{ax:probcontinuity}, for some $\alpha\in(0,1)$,
\[\alpha P^++(1-\alpha)R\succ P\succ\alpha P^-+(1-\alpha)R.\]
The outer lotteries have the same induced probability tree and support size at least four. Varying their prize vector and applying \Cref{ax:weakcontinuityinmoney:simple} therefore yields a large-support lottery $R_P\sim P$.

Define $V(P)=\sum_{z\in\supp R_P}v(R_P(z),z)-\bar c(R_P).$
This is well-defined because any two large-support lotteries indifferent to $P$ have the same represented value. Hence $V$ represents $\succsim$ on $\Delta(X)$.

\Cref{ax:probcontinuity} gives $V(\alpha P+(1-\alpha)Q)\to V(P)$ as $\alpha\uparrow1$, while \Cref{ax:weakcontinuityinmoney:simple} gives $V(P_n)\to V(P)$ whenever $P_n\to P$ on a fixed induced probability tree. For either claim, choose a large-support lottery $S\sim P$ and perturbations $S^+\succ P\succ S^-$ on the same induced probability tree whose represented values are arbitrarily close to $V(P)$; the corresponding axiom places the nearby lotteries between $S^+$ and $S^-$.

For every nondegenerate $P$, set
\[c(P)=\sum_{x\in\supp P}v(P(x),x)-V(P).\]
If $P$ and $Q$ have the same induced probability tree, choose a large-support lottery $R$ with support disjoint from $\supp P\cup\supp Q$, and set $P_\alpha=\alpha P+(1-\alpha)R$ and $Q_\alpha=\alpha Q+(1-\alpha)R$. Then $\bar c(P_\alpha)=\bar c(Q_\alpha)$. By \Cref{lem:smooth-prob-continuity} and the preceding continuity of $V$, $\bar c(P_\alpha)\to c(P)$ and $\bar c(Q_\alpha)\to c(Q)$, so $c(P)=c(Q)$. Moreover, $c(P)=\bar c(P)$ whenever $|\supp P|\geq4$.

It remains to extend $v$ to probability one. Normalize the large-support representation so that $V(\delta_{x^0})=0$. Choose a large-support lottery $R$ whose prizes differ from $x^0$, and let $P_\alpha^0=\alpha\delta_{x^0}+(1-\alpha)R$. Since $v(\alpha,x^0)=0$, $v((1-\alpha)R(z),z)\to0$, and $V(P_\alpha^0)\to0$, we have $\bar c(P_\alpha^0)\to0$.

Fix $x\in X$ and choose $R$ with support disjoint from $\{x,x^0\}$. Let $P_\alpha^x=\alpha\delta_x+(1-\alpha)R$. Since $P_\alpha^x$ and $P_\alpha^0$ have the same induced probability tree, $\bar c(P_\alpha^x)=\bar c(P_\alpha^0)\to0$, and therefore
\[v(\alpha,x)=V(P_\alpha^x)+\bar c(P_\alpha^x)-\sum_{z\in\supp R}v((1-\alpha)R(z),z)\longrightarrow I(\delta_x).\]
Set $v(1,x)=V(\delta_x)$ and $c(\delta_x)=0$. Then $v(\cdot,x)$ is continuous on $[0,1]$. Since $V$ represents $\succsim$, \Cref{ax:samedistfosd} implies that $v(1,\cdot)$ is strictly increasing, while its continuity follows from the within-tree continuity of $V$. Thus $c$ is a perturbation function and the stated representation holds on all of $\Delta(X)$.

Finally, let $P_\alpha=\alpha P+(1-\alpha)Q$. We already know that $V(P_\alpha)\to V(P)$. Since $v(\cdot,x)$ is continuous on $[0,1]$ and $v(0,x)=0$, the branch-utility sum of $P_\alpha$ converges to that of $P$. Hence $c(P_\alpha)\to c(P)$, so $c$ is linearly continuous.
\end{proof}

\begin{lemma}{lem:smooth-high-prob-fosd}
Under the notation of \Cref{lem:smooth-extension}, define $u(x)=4v(1/4,x)$ and, for $p\in(1/2,1]$, $d_p(x)=v(p,x)-v(1-p,x).$ Then the representation satisfies \Cref{ax:samedistfosd} iff $d_p$ is strictly increasing for every $p\in(1/2,1]$.
\end{lemma}

\begin{proof}
Suppose first that \Cref{ax:samedistfosd} holds. Fix $p\in(1/2,1)$ and $x>y$. Then $p\delta_x+(1-p)\delta_y$ strictly first-order stochastically dominates $p\delta_y+(1-p)\delta_x$. Since the two lotteries have the same induced probability tree,
\[v(p,x)+v(1-p,y)>v(p,y)+v(1-p,x),\]
so $d_p(x)>d_p(y)$. The case $p=1$ follows from $\delta_x\succ\delta_y$ whenever $x>y$.

Conversely, suppose that every $d_p$ is strictly increasing, and let $P=(r,x)$ strictly first-order stochastically dominate $Q=(r,y)$. If $r_i\leq1/2$ for every $i$, the representation on this tree is expected utility with utility $u$, so $P\succ Q$.

Otherwise, let $r_h=p>1/2$ be the unique high-probability branch. First-order stochastic dominance implies $x_h\geq y_h$. Define the finite measures $\widehat P=P-(2p-1)\delta_{x_h}$ and $\widehat Q=Q-(2p-1)\delta_{y_h}$. They have equal mass $2(1-p)$ and satisfy $F_{\widehat P}(t)\leq F_{\widehat Q}(t)$ for every $t$. For $t<y_h$ and $t\geq x_h$ this follows directly from the dominance of $P$ over $Q$; for $y_h\leq t<x_h$, $F_P(t)\leq1-p$ and $F_Q(t)\geq p$.

Hence $\int u\,d\widehat P\geq\int u\,d\widehat Q$. If $x_h>y_h$, strict monotonicity of $d_p$ gives $d_p(x_h)>d_p(y_h)$; if $x_h=y_h$, strict dominance of $P$ over $Q$ implies strict dominance of $\widehat P$ over $\widehat Q$. Thus
\[\int u\,d\widehat P+d_p(x_h)> \int u\,d\widehat Q+d_p(y_h).\]
Since $v(p,\cdot)=(1-p)u+d_p$, these are the branch-utility sums of $P$ and $Q$. Their perturbation terms coincide, so $P\succ Q$.
\end{proof}

\begin{proof}[Proof of \Cref{thm:variationalweightedfull}]
Suppose \Cref{ax:weakorder:simple,ax:weakcontinuityinmoney:simple,ax:probcontinuity,ax:samedistfosd,ax:weakcindependence} hold. \Cref{lem:smooth-small-prob-linearity,lem:smooth-prob-continuity,lem:smooth-extension,lem:smooth-high-prob-fosd} give the stated representation and its properties.

The necessity of the axioms is standard and omitted.
\end{proof}

\begin{proof}[Proof of \Cref{thm:variationalweighted}]
Suppose \Cref{ax:weakorder:simple,ax:weakcontinuityinmoney:simple,ax:probcontinuity,ax:samedistfosd,ax:weakcindependence,ax:splittinghalf} hold. By \Cref{thm:variationalweightedfull}, let
$u=4v(1/4,\cdot)$, so $v(r,\cdot)=ru$ for every $r\leq1/2$.

Fix $p\in(1/2,1)$ and let $q=1-p$. On the binary tree $(p,q)$, preferences are represented by $v(p,a)+qu(b)$. Choose $\alpha\in(0,1)$ with $\alpha p<1/2$. For binary lotteries $P=p\delta_a+q\delta_b$ and $Q=p\delta_c+q\delta_d$, choose $z\notin\supp P\cup\supp Q$. By \Cref{ax:splittinghalf}, $P\succsim Q$ implies $\alpha P+(1-\alpha)\delta_z\succsim\alpha Q+(1-\alpha)\delta_z$. The common branch and perturbation terms cancel, while $\alpha p,\alpha q<1/2$. Hence
\[v(p,a)+qu(b)\geq v(p,c)+qu(d) \quad\Longrightarrow\quad pu(a)+qu(b)\geq pu(c)+qu(d).\]

Equality on the right implies equality on the left. Indeed, if the right-hand expressions were equal but the left-hand inequality were strict, a sufficiently small downward perturbation of one prize on the preferred side would preserve the latter strict inequality while reversing the former, contradicting the implication above. The opposite strict ranking is symmetric. By continuity, the two additive expressions therefore represent the same weak order on $X^2$.

Uniqueness of additive representations gives $\eta>0$ and $\beta_1,\beta_2\in\R$ such that
\[v(p,\cdot)=\eta pu+\beta_1,\qquad qu=\eta qu+\beta_2.\]
Since $q>0$ and $u$ is nonconstant, $\eta=1$ and $\beta_2=0$; the normalization $v(p,x^0)=u(x^0)=0$ then gives $\beta_1=0$. Thus $v(p,\cdot)=pu$. Since $p\in(1/2,1)$ was arbitrary and $v(\cdot,x)$ is continuous at 1, define $V(P)=\sum_{x\in\supp P}P(x)u(x)-c(P).$ Then $V$ represents $\succsim$.

The necessity of the axioms is standard and omitted.
\end{proof}

\section{Counterexample for Theorem 1 of \textcite{puri2025}}\label{app:puricounterexample}

Let $\mu(P)=\sum_{x\in\supp P}P(x)x$ and $\ell(P)=\mathbf{1}(|\supp P|\geq 3)$. Define a lexicographic order $\succsim^\ast$ on $\Delta(X)$ by
\[P\succsim^\ast Q \quad\Longleftrightarrow\quad \ell(P)<\ell(Q) \ \text{or}\ \bigl[\ell(P)=\ell(Q)\ \text{and}\ \mu(P)\geq\mu(Q)\bigr].\]
Thus every lottery with at most two prizes is strictly preferred to every lottery with at least three prizes, while lotteries within either class are ranked by expected value. This is a complete and transitive preference.

$\succsim^\ast$ satisfies Puri's Axioms 1 to 5. For Axiom 1, if $P$ and $Q$ have certainty equivalents, then for every admissible $R$ and $\alpha\in(0,1)$ the two lotteries
\[\alpha\left(\frac12P+\frac12\delta_{\mu(Q)}\right)+(1-\alpha)R \quad\text{and}\quad \alpha\left(\frac12Q+\frac12\delta_{\mu(P)}\right)+(1-\alpha)R\]
have the same support and the same expected value, and hence are indifferent. Axiom 2 follows immediately from the expected-value ranking of certainties. Same-support independence and continuity hold because, on each class, preferences are represented by expected value. For Axiom 4, take $z=(x+y)/2$. For Axiom 5, take any $z<2y-x$. In both cases the lotteries being compared have the same three-prize support, so the required comparisons follow from expected value.

However, $\succsim^\ast$ does not admit a simplicity representation. Suppose instead that
\[V(P)=\sum_x P(x)u(x)-C(|\supp P|)\]
represented $\succsim^\ast$, where $u$ is continuous, strictly increasing, and unbounded below and $C$ is real-valued. Fix any three-prize lottery $P$. Since $V(P)\in\mathbb R$ and $u$ is unbounded below, there exists $y\in X$ such that $u(y)-C(1)<V(P)$. The representation would then imply $P\succ^\ast\delta_y$. By construction, however, $\delta_y\succ^\ast P$ for every $y$. This is a contradiction.

\section{Proof of \Cref{thm:costforsupport}}

\begin{lemma}{lem:support-size-solvability}
Suppose that \Cref{ax:weakorder:simple,ax:weakcontinuityinmoney:simple,ax:samedistfosd,ax:unbounded} hold. Let $r\in\Delta^4(\N)$, let $A\subseteq\supp r$ with $|A|\leq2$, and fix distinct prizes $(a_k)_{k\in A}$. Then, for every $\bar P\in\Delta(X)$, there exists $z\in\ndiag{r}$ such that $z_k=a_k$ for $k\in A$ and $(r,z)\sim\bar P$.
\end{lemma}

\begin{proof}
For any $P,Q\in\Delta(X)$, $\alpha\in[0,1)$, and finite $G\subset X$, \Cref{ax:unbounded} supplies $x^+,x^-\notin G$ such that
\[\alpha P+(1-\alpha)\delta_{x^+}\succ Q\succ\alpha P+(1-\alpha)\delta_{x^-}.\]
Indeed, first choose certainties above and below the finite set $\{Q\}\cup\{\alpha P+(1-\alpha)\delta_g:g\in G\}$, and then apply \Cref{ax:unbounded} with $P$ and $\alpha$.

Choose distinct $i,j\in\supp r\setminus A$, set $B=\supp r\setminus\{i,j\}$, and choose distinct $(b_k)_{k\in B}$ with $b_k=a_k$ for $k\in A$. Choose $t^0$ below all $b_k$. Applying the preceding observation to coordinate $i$, followed by \Cref{ax:samedistfosd}, gives $t^+$ above $t^0$ and all $b_k$ such that
\[r_i\delta_{t^+}+r_j\delta_{t^0}+\sum_{k\in B}r_k\delta_{b_k}\succ\bar P.\]
With $t^+$ fixed, the same argument applied to coordinate $j$ gives $t^-<t^0$ such that
\[\bar P\succ r_i\delta_{t^+}+r_j\delta_{t^-}+\sum_{k\in B}r_k\delta_{b_k}.\]

For $t\in[t^-,t^0]$, let $z_i(t)=t^+$, $z_j(t)=t$, and $z_k(t)=b_k$ for $k\in B$. Then $z(t)\in\ndiag{r}$, and \Cref{ax:weakcontinuityinmoney:simple} and connectedness give $t^*\in[t^-,t^0]$ such that $(r,z(t^*))\sim\bar P$.
\end{proof}

\begin{proposition}{prop:fixed-tree-additive}
Suppose that $\succsim$ satisfies \Cref{ax:weakorder:simple,ax:weakcontinuityinmoney:simple,ax:samedistfosd,ax:weakcindependence}. For every $p\in\Delta^4(\N)$, $\succsim_p$ admits an additive representation $(u_i^p)_{i\in\supp p}$.
\end{proposition}

\begin{proof}
This follows from \textcite{mononen2026preference}.
\end{proof}

\begin{lemma}{lem:shared-probability-alignment}
Suppose that $\succsim$ satisfies the assumptions of \Cref{prop:fixed-tree-additive} and \Cref{ax:unbounded}. For each $p\in\Delta^4(\N)$, let $(u_i^p)_{i\in\supp p}$ be an additive representation of $\succsim_p$. Let $p,q\in\Delta^4(\N)$ and let distinct $i,j\in\supp p\cap\supp q$ satisfy $p_i=q_i$ and $p_j=q_j$. Then there exist $\eta>0$ and $\beta_i,\beta_j\in\R$ such that
\[u_i^q=\eta u_i^p+\beta_i,\qquad u_j^q=\eta u_j^p+\beta_j.\]
\end{lemma}

\begin{proof}
Choose $x\in\ndiag{p}$. By \Cref{lem:support-size-solvability}, there exists $y\in\ndiag{q}$ such that $y_i=x_i$, $y_j=x_j$, and $(p,x)\sim(q,y)$.

For any admissible pair $(a,b)$, replace coordinates $i,j$ of both prize vectors by $a,b$. Two applications of \Cref{ax:weakcindependence} preserve the indifference between the resulting lotteries. Hence, for any admissible pairs $(a,b)$ and $(c,d)$,
\[u_i^p(a)+u_j^p(b)\geq u_i^p(c)+u_j^p(d) \iff u_i^q(a)+u_j^q(b)\geq u_i^q(c)+u_j^q(d).\]
The admissible pairs are dense in $X^2$, so continuity extends the equivalence to $X^2$. Uniqueness of additive representations \parencite[Theorem 13]{krantz1971} gives the result.
\end{proof}

\begin{proposition}{prop:common-branch-representation}
Suppose that $\succsim$ satisfies \Cref{ax:weakorder:simple,ax:weakcontinuityinmoney:simple,ax:samedistfosd,ax:weakcindependence,ax:unbounded}. Then there exists $w:(0,1)\times X\to\R$ such that $w(a,\cdot)$ is continuous and strictly increasing for every $a\in(0,1)$ and, for every $p\in\Delta^4(\N)$, $x\mapsto\sum_{i\in\supp p}w(p_i,x_i)$ represents $\succsim_p$.
\end{proposition}

\begin{proof}
For each $p\in\Delta^4(\N)$, choose an additive representation $(u_i^p)_{i\in\supp p}$ from \Cref{prop:fixed-tree-additive}. Fix $x^0<x^1$ in $X$ and normalize $u_i^p(x^0)=0$ for every $p$ and $i$.

On the uniform four-branch tree, symmetry and uniqueness imply that the four normalized coordinate indices coincide. Normalize their common scale at $x^1$ to one and call the resulting index $w(1/4,\cdot)$.

For $a\in(0,1/2)$, let
\[\theta(a)=(a,1/4,1/4,1/2-a).\]
For $a\neq1/4$, \Cref{lem:shared-probability-alignment} allows its representation to be normalized so that the $1/4$-coordinate indices equal $w(1/4,\cdot)$; let $w(a,\cdot)$ be its $a$-coordinate index. For $a=1/4$, $\theta(a)$ is the uniform four-branch tree. Once one shared coordinate has the same normalized index on two trees, \Cref{lem:shared-probability-alignment} fixes their positive affine transformation.

Fix $p\in\Delta^4(\N)$ and let $m=\min_i p_i\leq1/4$. If $m=1/4$, then $p$ is the uniform four-branch tree. Otherwise choose $i$ with $p_i=m$ and $j\neq i$ with $p_j$ minimal among the remaining probabilities. Since $p_j\leq(1-m)/3$, there exists a four-branch tree containing $m$, $p_j$, and $1/4$. Comparing it with $\theta(m)$ and $\theta(p_j)$, and then with $p$, aligns the $m$-coordinate of $p$ with $w(m,\cdot)$.

Now let $a=p_i<1/2$, $a\neq m$. Since $m\leq(1-a)/3$, there exists a four-branch tree containing $a$, $m$, and $1/4$. Comparing it with $\theta(a)$ and $\theta(m)$ aligns its $a$- and $m$-coordinate indices with $w(a,\cdot)$ and $w(m,\cdot)$. Comparing it with $p$ through $a$ and $m$ gives $u_i^p=w(a,\cdot)$. Hence every branch probability below $1/2$ has the corresponding index $w$.

Finally fix $a\in[1/2,1)$ and set $b=(1-a)/3$. On the tree $(a,b,b,b)$, the three $b$-coordinate indices are already $w(b,\cdot)$; let $w(a,\cdot)$ be the remaining coordinate index. If $p_i=a$, let $m=\min_{j\neq i}p_j$. Then $m\leq b$, so $(a,m,b,1-a-m-b)\in\Delta^4(\N)$. Comparing this tree with $(a,b,b,b)$ through $a,b$, and with $p$ through $a,m$, gives $u_i^p=w(a,\cdot)$.

Hence $x\mapsto\sum_{i\in\supp p}w(p_i,x_i)$ represents $\succsim_p$ for every $p\in\Delta^4(\N)$.
\end{proof}

\begin{lemma}{lem:local-cross-tree-representation}
Suppose that the assumptions of \Cref{prop:common-branch-representation} hold. Let $p,q\in\Delta^4(\N)$, $x\in\ndiag{p}$, and $y\in\ndiag{q}$ satisfy $(p,x)\sim(q,y)$. Then there exist $P^+,P^-\in\Delta(X)$ with $P^+\succ(p,x)\succ P^-$ such that, for all $\tilde x\in\ndiag{p}$ and $\tilde y\in\ndiag{q}$ satisfying $P^+\succsim(p,\tilde x)\succsim P^-$ and $P^+\succsim(q,\tilde y)\succsim P^-$,
\[(p,\tilde x)\succsim(q,\tilde y)\iff \sum_{i\in\supp p}\bigl[w(p_i,\tilde x_i)-w(p_i,x_i)\bigr] \geq \sum_{j\in\supp q}\bigl[w(q_j,\tilde y_j)-w(q_j,y_j)\bigr].\]
\end{lemma}

\begin{proof}
First consider two indifferent lotteries $(p,x)\sim(q,y)$ with $p_i=q_j$ and $x_i=y_j$ for some branches $i$ and $j$. Choose $z^-<x_i<z^+$ such that $[z^-,z^+]$ contains no other coordinate of $x$ or $y$, and let $x(z)$ and $y(z)$ replace coordinates $i$ and $j$, respectively, by $z$. By \Cref{ax:weakcindependence}, $(p,x(z))\sim(q,y(z))$ for every $z\in[z^-,z^+]$.

If $\hat x\in\ndiag{p}$ and $\hat y\in\ndiag{q}$ satisfy
$(p,x(z^+))\succsim(p,\hat x)\succsim(p,x(z^-))$ and
$(q,y(z^+))\succsim(q,\hat y)\succsim(q,y(z^-))$, continuity and strict monotonicity of $w(p_i,\cdot)$ give $s,t\in[z^-,z^+]$ such that $(p,\hat x)\sim(p,x(s))$ and $(q,\hat y)\sim(q,y(t))$. Since $(p,x(z))\sim(q,y(z))$ for every $z\in[z^-,z^+]$,
\[(p,\hat x)\succsim(q,\hat y)\iff w(p_i,s)\geq w(p_i,t).\]
The fixed-tree representations give
\[\sum_{k\in\supp p}\bigl[w(p_k,\hat x_k)-w(p_k,x_k)\bigr] =w(p_i,s)-w(p_i,x_i)\]
\[\text{and } \sum_{k\in\supp q}\bigl[w(q_k,\hat y_k)-w(q_k,y_k)\bigr] =w(p_i,t)-w(p_i,x_i).\]
Hence
\[(p,\hat x)\succsim(q,\hat y)\iff \sum_{k\in\supp p}\bigl[w(p_k,\hat x_k)-w(p_k,x_k)\bigr] \geq \sum_{k\in\supp q}\bigl[w(q_k,\hat y_k)-w(q_k,y_k)\bigr].\]

Now return to arbitrary $p,q,x,y$ in the statement. Choose $i\in\supp p$ with minimal probability, so $p_i\leq1/4$. Since the coordinates of $y$ are distinct and $|\supp q|\geq4$, at least three differ from $x_i$; among them choose $j$ with minimal probability. Then $q_j\leq1/3$, so $p_i+q_j<1$. Choose $r\in\Delta^4(\N)$ with $r_1=p_i$ and $r_2=q_j$.

By \Cref{lem:support-size-solvability}, choose $z\in\ndiag{r}$ such that $z_1=x_i$, $z_2=y_j$, and $(r,z)\sim(p,x)\sim(q,y)$. The preceding local calibration applies to $p$ and $r$ through branches $i,1$, and to $r$ and $q$ through branches $2,j$. Choose $P^+\succ(r,z)\succ P^-$ on the $r$-tree sufficiently close to $(r,z)$ that both calibrations apply between these preference levels.

Fix $\tilde x,\tilde y$ satisfying the stated bounds. By \Cref{lem:support-size-solvability}, choose $\tilde z\in\ndiag{r}$ with $(r,\tilde z)\sim(p,\tilde x)$. The calibration between $p$ and $r$ gives
\[\sum_{k\in\supp p}\bigl[w(p_k,\tilde x_k)-w(p_k,x_k)\bigr] = \sum_{k\in\supp r}\bigl[w(r_k,\tilde z_k)-w(r_k,z_k)\bigr].\]
The calibration between $r$ and $q$ gives
\[(p,\tilde x)\succsim(q,\tilde y) \iff \sum_{k\in\supp r}\bigl[w(r_k,\tilde z_k)-w(r_k,z_k)\bigr] \geq \sum_{k\in\supp q}\bigl[w(q_k,\tilde y_k)-w(q_k,y_k)\bigr].\]
\end{proof}

\begin{lemma}{lem:global-cross-tree-representation}
Suppose that the assumptions of \Cref{lem:local-cross-tree-representation} hold. Then there exists a function $K$ on induced probability trees with at least four branches such that, for all $p,q\in\Delta^4(\N)$, $x\in\ndiag{p}$, and $y\in\ndiag{q}$,
\[(p,x)\succsim(q,y)\iff \sum_{i\in\supp p}w(p_i,x_i)+K(p) \geq \sum_{j\in\supp q}w(q_j,y_j)+K(q).\]
\end{lemma}

\begin{proof}
By \Cref{prop:common-branch-representation}, $x\mapsto\sum_{i\in\supp p}w(p_i,x_i)$ is a continuous representation of $\succsim_p$ for every $p\in\Delta^4(\N)$. \Cref{lem:support-size-solvability} represents every preference level on every fixed tree, while \Cref{ax:samedistfosd} and openness of $X$ provide strictly better and worse lotteries on that tree around every lottery. \Cref{lem:local-cross-tree-representation} locally calibrates differences in the fixed-tree representations across trees. The local-to-global representation theorem of \textcite{mononen2026preference} therefore yields a function $K$ such that the displayed expression represents $\succsim$ globally.
\end{proof}

\begin{proposition}{prop:large-support-support-size}
Suppose that the assumptions of \Cref{lem:global-cross-tree-representation} and \Cref{ax:samesupportind} hold. Then there exist a continuous and strictly increasing function $u:X\to\R$ with $u(X)=\R$ and a function $\kappa:\{4,5,\ldots\}\to\R$ such that, for all lotteries $P,Q$ with $|\supp P|,|\supp Q|\geq4$,
\[P\succsim Q\iff\sum_{x\in\supp P}P(x)u(x)-\kappa(|\supp P|)\geq\sum_{x\in\supp Q}Q(x)u(x)-\kappa(|\supp Q|).\]
\end{proposition}

\begin{proof}
By \Cref{lem:global-cross-tree-representation}, a lottery $(p,x)$ is represented by $\sum_iw(p_i,x_i)+K(p)$. Fix $x^0\in X$. Subtracting $w(a,x^0)$ from $w(a,\cdot)$ and absorbing the resulting tree-specific constant into $K$, assume $w(a,x^0)=0$ for every $a\in(0,1)$.

For probabilities below $1/2$, the argument of \Cref{lem:smooth-small-prob-linearity} gives additivity in the probability argument. The monotone Cauchy argument therefore yields $w(a,x)=au(x)$ for every $a\in(0,1/2)$, where $u(x)=4w(1/4,x)$.

Fix $a\in[1/2,1)$ and let $b=(1-a)/4$. For distinct prizes $x,y,z,v$, compare the lotteries with probability vectors $(a,b,2b,b)$ and $(b,a,b,2b)$ on the common support $\{x,y,z,v\}$. Their $K$-terms coincide, and the comparison is equivalent to
\[w(a,x)-bu(x)+bu(z)\geq w(a,y)-bu(y)+bu(v).\]
Mix both lotteries with the uniform lottery on the same support. For sufficiently small $\alpha>0$, every resulting branch probability is below $1/2$. By \Cref{ax:samesupportind}, the original comparison is therefore equivalent to the expected-utility comparison
\[(a-b)u(x)+bu(z)\geq(a-b)u(y)+bu(v).\]
Density and continuity extend this equivalence to $X^2$. The two additive representations have the same nonconstant second-coordinate index $bu$, so uniqueness of additive representations fixes their common scale and implies that $w(a,\cdot)-bu$ and $(a-b)u$ differ only by a constant. Both vanish at $x^0$, so the constant is zero. Hence $w(a,\cdot)=au$. Thus $w(a,x)=au(x)$ for every $a\in(0,1)$, and every large-support lottery $(p,x)$ is represented by $\sum_i p_i u(x_i)+K(p)$.

Suppose $\sup u(X)<\infty$. Fix a large-support lottery $P$ and $\alpha\in(0,1)$. As $x\notin\supp P$ varies, the lotteries $\alpha P+(1-\alpha)\delta_x$ have the same induced probability tree, and their expected-utility terms are bounded above by $\alpha\sum_{z\in\supp P}P(z)u(z)+(1-\alpha)\sup u(X)<\sup u(X)$. Choose a lottery $Q$ with this induced probability tree and expected utility above this bound. Since the $K$-terms coincide, $Q\succ\alpha P+(1-\alpha)\delta_x$ for every $x\notin\supp P$. The finite-avoidance argument in the proof of \Cref{lem:support-size-solvability} gives $x\notin\supp P$ such that $\alpha P+(1-\alpha)\delta_x\succ Q$, a contradiction. Thus $u$ is unbounded above. The argument for unboundedness below is symmetric, so $u(X)=\R$.

Fix $n\geq4$, distinct strictly positive $n$-dimensional probability vectors $p,q$, and $\alpha\in(0,1)$. For distinct prizes $x_1,\ldots,x_n$, \Cref{ax:samesupportind}, applied with the $q$-lottery as the common mixture, gives
\[\sum_i(p_i-q_i)u(x_i)+K(p)-K(q)\geq0 \iff \alpha\sum_i(p_i-q_i)u(x_i)+K\bigl(\alpha p+(1-\alpha)q\bigr)-K(q)\geq0.\]
Since $u(X)=\R$, the expected-utility difference ranges densely over $\R$ as the distinct prizes vary. The two inequalities therefore have the same cutoff, which gives $K(\alpha p+(1-\alpha)q)=\alpha K(p)+(1-\alpha)K(q)$. Hence $K$ is affine on the interior of each $n$-simplex.

Because $K$ depends only on the induced probability tree, it is permutation invariant. The uniform probability vector is the average of the permutations of any $p$, so affinity and permutation invariance imply $K(p)=K(1/n,\ldots,1/n)$. Thus $K$ is constant over all $n$-branch induced probability trees. Write this common value as $-\kappa(n)$. This gives the stated representation.
\end{proof}

For each nondegenerate lottery $P$, choose $R_P\sim P$ with $|\supp R_P|\geq4$, as guaranteed by \Cref{lem:support-size-solvability}, taking $R_P=P$ when $|\supp P|\geq4$, and set
\[\widehat V(P)=\sum_{x\in\supp R_P}R_P(x)u(x)-\kappa(|\supp R_P|).\]
This is well defined and represents $\succsim$ on the nondegenerate domain.

\begin{lemma}{lem:small-support-calibration}
The function $\kappa$ from \Cref{prop:large-support-support-size} admits an extension to $\{2,3,\ldots\}$ such that every nondegenerate lottery $P$ satisfies $\widehat V(P)=\sum_{x\in\supp P}P(x)u(x)-\kappa(|\supp P|)$.
\end{lemma}

\begin{proof}
Fix $n\in\{2,3\}$ and suppose the representation is established for support size $n+1$. Fix a strictly positive probability vector $p=(p_1,\ldots,p_n)$, distinct prizes $x_1,\ldots,x_n,y$, and let $\alpha=1-p_n$. Set $P=\alpha \sum_{i<n}(p_i/\alpha)\delta_{x_i}+p_n\delta_{x_n}$ and $P'=\alpha \sum_{i<n}(p_i/\alpha)\delta_{x_i}+p_n\delta_y$.

For every $t\in\R$, choose an $n$-point lottery $Q$ with support disjoint from $\{x_1,\ldots,x_n,y\}$ and $\sum_zQ(z)u(z)=t$, which is possible because $u(X)=\R$. By \Cref{ax:weakcindependence},
\[P\succsim\alpha Q+p_n\delta_{x_n}\iff P'\succsim\alpha Q+p_n\delta_y.\]
The two comparison lotteries have support size $n+1$, so
\[\widehat V(P)-p_nu(x_n)\geq\alpha t-\kappa(n+1) \iff \widehat V(P')-p_nu(y)\geq\alpha t-\kappa(n+1).\]
Since $t$ ranges over $\R$, the two cutoffs coincide. Hence replacing one prize leaves $\widehat V(P)-\sum_xP(x)u(x)$ unchanged. Repeating the argument coordinatewise shows that, for each fixed probability vector $p$, this difference is independent of the prizes.

Now let $p$ and $q$ be strictly positive $n$-dimensional probability vectors and assign both to the same distinct prizes $x_1,\ldots,x_n$. Same-Support Independence and \Cref{prop:large-support-support-size}, after mixing both lotteries with a common lottery whose support is disjoint and makes the mixtures have at least four prizes, give
\[\sum_i p_i u(x_i)\geq\sum_i q_i u(x_i) \iff \widehat V\!\left(\sum_i p_i\delta_{x_i}\right)\geq \widehat V\!\left(\sum_i q_i\delta_{x_i}\right).\]
The difference between $\widehat V$ and expected utility is already independent of the prizes for each probability vector. Since $u(X)=\R$, the expected-utility difference between the two lotteries can approach zero from either side while the prizes remain distinct. The displayed rankings can therefore agree for every such prize vector only if the two differences between $\widehat V$ and expected utility coincide. Hence this difference depends only on $n$; write its common value as $-\kappa(n)$.

Apply the argument first with $n=3$, using \Cref{prop:large-support-support-size}, and then with $n=2$.
\end{proof}

Every lottery has a certainty equivalent: \Cref{ax:unbounded} with $\alpha=0$ brackets it by certainties, and \Cref{ax:weakcontinuityinmoney:simple} and connectedness of $X$ give an indifferent certainty.

\begin{lemma}{lem:certainty-calibration}
Let $u$ and $\widehat V$ be given by the preceding nondegenerate representation. There exists a continuous and strictly increasing function $\phi:\R\to\R$ with $\phi(\R)=\R$ such that, for every $x\in X$ and every nondegenerate lottery $P$, $\delta_x\succsim P\iff\phi(u(x))\geq\widehat V(P)$.
\end{lemma}

\begin{proof}
For each $x\in X$, define $\phi(u(x))$ as the $\widehat V$-value of any nondegenerate lottery indifferent to $\delta_x$, whose existence follows from \Cref{lem:support-size-solvability}. This is well defined because $\widehat V$ represents $\succsim$ on the nondegenerate domain. Hence $\delta_x\succsim P$ if and only if $\phi(u(x))\geq\widehat V(P)$.

If $x>y$, then $\delta_x\succ\delta_y$ by \Cref{ax:samedistfosd}, so $\phi(u(x))>\phi(u(y))$. Since $u$ is strictly increasing and $u(X)=\R$, $\phi$ is strictly increasing.

For continuity, suppose that $\phi$ has a left jump at $u(x)$. Since $\widehat V$ has range $\R$ on the nondegenerate domain, choose a nondegenerate lottery $P$ with
\[\sup_{y<x}\phi(u(y))<\widehat V(P)<\phi(u(x)).\]
For any sequence $x_n\uparrow x$, $\delta_x\succ P\succ\delta_{x_n}$, contradicting \Cref{ax:probcontinuity}. A right jump is ruled out symmetrically, so $\phi$ is continuous.

By \Cref{lem:small-support-calibration} and $u(X)=\R$, $\widehat V$ has range $\R$ on the nondegenerate domain.
\end{proof}

\begin{proof}[Proof of \Cref{thm:costforsupport}]
Suppose \Cref{ax:weakorder:simple,ax:weakcontinuityinmoney:simple,ax:samedistfosd,ax:weakcindependence,ax:unbounded,ax:samesupportind} hold. By \Cref{prop:large-support-support-size,lem:small-support-calibration}, every nondegenerate lottery $P$ is represented by $\widehat V(P)=\sum_{x\in\supp P}P(x)u(x)-\kappa(|\supp P|)$. By \Cref{lem:certainty-calibration}, certainty is represented on the same scale by $\phi(u(x))$. Hence the function $V$ in \Cref{thm:costforsupport} represents $\succsim$ on $\Delta(X)$.

Conversely, the representation satisfies Weak Order and Within-Tree Prize Continuity by continuity, and \Cref{ax:samedistfosd} by strict monotonicity of $u$ and $\phi\circ u$. Branch Cancellation follows because replacing the common prize leaves the difference between the two augmented values unchanged. Unboundedness follows from $u(X)=\phi(\R)=\R$. Finally, if $P$ and $Q$ have the same support, then their mixtures with any common $R$ have the same support and their represented-value difference is $\alpha$ times the original difference, giving \Cref{ax:samesupportind}.
\end{proof}

\begin{proof}[Proof of \Cref{prop:costforsupportfull}]
Suppose the axioms in \Cref{prop:costforsupportfull} hold. By \Cref{thm:costforsupport}, let $u,\phi,\kappa$ represent $\succsim$. Fix $x,y\in X$. Since $u(X)=\R$, there exist binary lotteries $P,Q$ with a common support disjoint from $\{x,y\}$ such that $P\sim\delta_x$ and $Q\sim\delta_y$. By \Cref{ax:certaintyequivalentswap},
\[\tfrac12P+\tfrac12\delta_y\sim\tfrac12Q+\tfrac12\delta_x.\]
The two mixtures have the same support size, so their support size terms cancel and $\phi(u(x))+u(y)=\phi(u(y))+u(x)$. Since $u(X)=\R$, $\phi(t)=t+C$ for some $C\in\R$. Subtracting $C$ from the representing index and adding $C$ to $\kappa(n)$ for $n\geq2$ yields the representation in \Cref{prop:costforsupportfull} with $\kappa(1)=0$.

Conversely, this representation is the representation in \Cref{thm:costforsupport} with $\phi(t)=t$, so the axioms of that theorem follow from its converse. If $P\sim\delta_x$ and $Q\sim\delta_y$ have the same support, then their representation gives $\sum_zP(z)u(z)+u(y)=\sum_zQ(z)u(z)+u(x)$. The two swapped mixtures have the same support size and are therefore indifferent, establishing \Cref{ax:certaintyequivalentswap}.
\end{proof}

\section{Proofs of \Cref{thm:concave-information,cor:monotone-support-size-costs}}
\label{app:simplicity-proof}
\begin{proof}[Proof of \Cref{thm:concave-information}]
If $c$ is concave, affinity of expected utility gives
\[V(\alpha P+(1-\alpha)Q)\leq\alpha V(P)+(1-\alpha)V(Q)\leq\max\{V(P),V(Q)\},\]
so \Cref{ax:convexity} follows.

Conversely, fix distinct $P,Q\in\Delta(X)$ and $\alpha\in(0,1)$. Since the coefficients $P(x)-Q(x)$ sum to zero and are not all zero and $u(X)=\R$, common injective relabelings of the prizes can make the expected-utility difference between $P$ and $Q$ converge to $c(P)-c(Q)$. Let $P_n,Q_n$ denote the resulting lotteries. Tree invariance gives $c(P_n)=c(P)$, $c(Q_n)=c(Q)$, and
$c\bigl(\alpha P_n+(1-\alpha)Q_n\bigr)=c\bigl(\alpha P+(1-\alpha)Q\bigr),$ while $V(P_n)-V(Q_n)\to0$. By \Cref{ax:convexity}, $V\bigl(\alpha P_n+(1-\alpha)Q_n\bigr)\leq\max\{V(P_n),V(Q_n)\}.$ Affinity of expected utility therefore gives
\[\alpha c(P)+(1-\alpha)c(Q)-c\bigl(\alpha P+(1-\alpha)Q\bigr)\leq|V(P_n)-V(Q_n)|.\]
Taking $n\to\infty$ proves concavity. The case $P=Q$ is immediate.
\end{proof}

Concavity and $c(\delta_x)=0$ give $c(P)\geq0$. For the branch-splitting implication, let $S$ be the split lottery and let $P_x,P_y$ be the two lotteries obtained by assigning the combined split probability entirely to either split prize. Then $S$ is a mixture of $P_x$ and $P_y$, while tree invariance gives $c(P_x)=c(P_y)=c(P)$. Hence concavity gives $c(S)\geq c(P)$.

\begin{proof}[Proof of \Cref{cor:monotone-support-size-costs}]
Suppose $\kappa$ is weakly increasing. Since $\supp\bigl(\alpha P+(1-\alpha)Q\bigr)=\supp P\cup\supp Q$,
\[\kappa\bigl(|\supp(\alpha P+(1-\alpha)Q)|\bigr)\geq\alpha\kappa(|\supp P|)+(1-\alpha)\kappa(|\supp Q|).\]
Affinity of expected utility then gives
\[V(\alpha P+(1-\alpha)Q)\leq\alpha V(P)+(1-\alpha)V(Q)\leq\max\{V(P),V(Q)\},\]
which implies \Cref{ax:convexity}.

Conversely, for $x>y$, \Cref{ax:convexity} gives $\delta_x\succsim\alpha\delta_x+(1-\alpha)\delta_y$, and hence $\kappa(2)\geq-(1-\alpha)(u(x)-u(y))$. Letting $\alpha\uparrow1$ yields $\kappa(2)\geq0=\kappa(1)$.

For $n\geq2$, choose $x_0<\cdots<x_n$ and full-support lotteries $P$ on $\{x_1,\ldots,x_n\}$ and $Q$ on $\{x_0,x_2,\ldots,x_n\}$ with the same expected utility. Such lotteries exist because the attainable expected-utility intervals overlap. Then $P\sim Q$, while $\alpha P+(1-\alpha)Q$ has support size $n+1$ and the same expected utility. By \Cref{ax:convexity}, $P\succsim\alpha P+(1-\alpha)Q$, so $\kappa(n+1)\geq\kappa(n)$. Hence $\kappa$ is weakly increasing.
\end{proof}

\section{Proof of \Cref{thm:separablevariationalweighted}}

\begin{proof}[Proof of \Cref{thm:separablevariationalweighted}]
Suppose that \Cref{ax:weakorder:simple,ax:weakcontinuityinmoney:simple,ax:probcontinuity,ax:lotteryweakcindependence,ax:samedistfosd,ax:splittinghalf} hold. Since \Cref{ax:lotteryweakcindependence} implies \Cref{ax:weakcindependence}, \Cref{thm:variationalweighted} yields a representation
\[V(P)=\sum_{x}P(x)u(x)-c(P),\]
where $u$ is continuous and strictly increasing and $c$ is a linearly
continuous perturbation function. Linear continuity implies that
$t\mapsto c(tA+(1-t)B)$ is continuous on $[0,1]$ for all lotteries
$A,B$.

\smallskip
\noindent\textit{Step 1: Cost differences are independent of the background.}
Fix $\alpha\in(0,1)$ and, for any $R$ with
$\supp R\cap(\supp P\cup\supp Q)
=\emptyset$, write
\[D_R(P,Q)=c\bigl(\alpha P+(1-\alpha)R\bigr) -c\bigl(\alpha Q+(1-\alpha)R\bigr).\]
Because $c$ depends only on the induced probability tree, $D_R(P,Q)$
depends only on the trees of $P,Q,R$. Moreover,
\begin{equation}\label{eq:prop4-background-comparison} \alpha P+(1-\alpha)R\succsim\alpha Q+(1-\alpha)R \iff \alpha\bigl(\sum_{x}P(x)u(x)-\sum_{x}Q(x)u(x)\bigr)\geq D_R(P,Q).\end{equation}
Choose $a<b$ in $X$ and let
$\varepsilon=\alpha\bigl(u(b)-u(a)\bigr)>0$.
For any fixed probability trees of $P$ and $Q$, assigning distinct
prizes in $(a,b)$ while avoiding a prescribed finite set makes
$\alpha(\sum_{x}P(x)u(x)-\sum_{x}Q(x)u(x))$ dense in $(-\varepsilon,\varepsilon)$.
Suppose $|D_R(P,Q)|<\varepsilon$ but
$D_S(P,Q)\neq D_R(P,Q)$ for another lottery $S$. Let
$G=\supp R\cup\supp S$,
and write
$P=\sum_{i=1}^m p_i\delta_{x_i}$ and
$Q=\sum_{j=1}^n q_j\delta_{y_j}$.

For any $t\in(-\varepsilon,\varepsilon)$, choose
$v,w\in(u(a),u(b))$ such that $\alpha(v-w)=t$.
Since $G$ is finite, we can choose sequences of
pairwise distinct prizes
$x_i^{(k)},y_j^{(k)}\in(a,b)\setminus G$
such that $u(x_i^{(k)})\to v$ and
$u(y_j^{(k)})\to w$ for every $i,j$.
Defining
\[P^{(k)}=\sum_{i=1}^m p_i\delta_{x_i^{(k)}}, \quad\text{and}\quad Q^{(k)}=\sum_{j=1}^n q_j\delta_{y_j^{(k)}},\]
we obtain $\alpha\bigl(\sum_{x}P^{(k)}(x)u(x)-\sum_{x}Q^{(k)}(x)u(x)\bigr)
    \rightarrow t$.
Moreover, since the probability trees are unchanged
and the new prizes avoid $G$,
\[D_R(P^{(k)},Q^{(k)})=D_R(P,Q)\quad \text{and}\quad D_S(P^{(k)},Q^{(k)})=D_S(P,Q)\]

Suppose first that $D_R(P,Q)<D_S(P,Q)$.
Since $|D_R(P,Q)|<\varepsilon$, we can choose
\[t\in\bigl(D_R(P,Q), \min\{D_S(P,Q),\varepsilon\}\bigr).\]
For sufficiently large $k$,
\[D_R(P,Q) <\alpha\bigl(\sum_{x}P^{(k)}(x)u(x)-\sum_{x}Q^{(k)}(x)u(x)\bigr) <D_S(P,Q).\]
By \eqref{eq:prop4-background-comparison},
the comparison under $R$ holds, whereas the
comparison under $S$ fails, contradicting \Cref{ax:lotteryweakcindependence}. If $D_S(P,Q)<D_R(P,Q)$,
the same argument applies with
\[t\in\bigl(\max\{D_S(P,Q),-\varepsilon\}, D_R(P,Q)\bigr).\]
Consequently,
\begin{equation}\label{eq:prop4-small-differences} |D_R(P,Q)|<\varepsilon \quad\Longrightarrow\quad D_R(P,Q)=D_S(P,Q).\end{equation}

For arbitrary $P,Q$, let $P_t=tP+(1-t)Q$. The function
\[t\longmapsto c\bigl(\alpha P_t+(1-\alpha)R\bigr)\]
is uniformly continuous on $[0,1]$. Choose a finite partition
$0=t_0<\cdots<t_N=1$ such that
$|D_R(P_{t_i},P_{t_{i-1}})|<\varepsilon$ for every $i$.
Both $R$ and $S$ remain admissible for each pair because
$\supp P_t\subseteq
\supp P\cup\supp Q$.
Applying \eqref{eq:prop4-small-differences} to each consecutive pair
and summing yields, by telescoping, $D_R(P,Q)=D_S(P,Q).$

\smallskip
\noindent\textit{Step 2: Branch splits satisfy the cocycle identity.}
For $x,y>0$ with $x+y\leq 1$, define
\[\theta(x,y) =c\bigl(x\delta_a+y\delta_b+(1-x-y)R\bigr) -c\bigl((x+y)\delta_a+(1-x-y)R\bigr),\]
where $a,b$ are distinct and outside $\supp R$.
When $x+y<1$, by step 1 and tree
invariance show that $\theta$ does not depend on $R,a,b$; when
$x+y=1$, the expression equals $c(x\delta_a+y\delta_b)$ and is
likewise well defined. Set $\theta(x,0)=\theta(0,x)=0$.

Tree invariance gives $\theta(x,y)=\theta(y,x)$. Merging three
branches in either order gives
\[\theta(x,y)+\theta(x+y,z) =\theta(y,z)+\theta(x,y+z) \qquad (x,y,z\geq0,\ x+y+z\leq1).\]
For fixed distinct $a,b,d\in X$,
\[\theta(x,h) =c\bigl(x\delta_a+h\delta_b+(1-x-h)\delta_d\bigr) -c\bigl((x+h)\delta_a+(1-x-h)\delta_d\bigr).\]
Continuity of $c$ along mixture segments implies that
$x\mapsto\theta(x,h)$ is continuous on $(0,1-h)$ and
$\theta(x,h)\to0$ as $h\downarrow0$ for each fixed $x\in(0,1)$.
Likewise,
\[\theta(h,1-h)=c\bigl(h\delta_a+(1-h)\delta_b\bigr) \longrightarrow c(\delta_b)=0.\]

\smallskip
\noindent\textit{Step 3: Recover a continuous additive perturbation.}
By \textcite{ng1974} (see also \textcite{ebanks1981}),
the symmetric cocycle equation on the closed triangle
implies that there exists $F:[0,1]\to\R$ such that
\[\theta(x,y)=F(x)+F(y)-F(x+y) \qquad (x,y\geq0,\ x+y\leq1).\]
In particular, $F(0)=0$. For every $h\in(0,1)$,
\[F(x+h)-F(x)=F(h)-\theta(x,h)\]
is continuous in $x\in(0,1-h)$. By \textcite[Theorem~1.3]{deBruijn1951},
there is an additive function $A:\R\to\R$
such that $\psi=F-A$ is continuous on $(0,1)$.
Since $F(0)=A(0)=0$ and $A$ is additive, we have
$\psi(0)=0$ and
\[\theta(x,y)=\psi(x)+\psi(y)-\psi(x+y) \qquad \text{for }x,y\geq0,\ x+y\leq1.\]
For fixed $x\in(0,1)$, the limits in Step~2 give
\begin{align*}\psi(h)&=\theta(x,h)+\psi(x+h)-\psi(x) \xrightarrow{h\to0}0\\ \text{and }\qquad \psi(1-h)&=\psi(1)+\theta(h,1-h)-\psi(h) \xrightarrow{h\to0}\psi(1).\end{align*}
Thus $\psi$ is continuous on $[0,1]$ with $\psi(0)=0$.

Let $P=\sum_{i=1}^n p_i\delta_{x_i}$ have distinct prizes, and write
$s_k=\sum_{i=1}^k p_i$. Successively merging the branches at \(x_2,\ldots,x_n\) into the branch at $x_1$ and using
$c(\delta_{x_1})=0$ gives, for $n\geq2$,
\[c(P)=\sum_{k=2}^n\theta(s_{k-1},p_k) =\sum_{k=2}^n \bigl[\psi(s_{k-1})+\psi(p_k)-\psi(s_k)\bigr] =\sum_{i=1}^n\psi(p_i)-\psi(1).\]
The formula also holds for $n=1$. Therefore
\[V(P)=\sum_{x}P(x)u(x) -\sum_{i=1}^n\psi(p_i)+\psi(1),\]
as required.

The necessity of the axioms is standard and omitted.
\end{proof}

\end{document}